%% file: main.tex
\documentclass[%
 reprint,
 amsmath,amssymb,
 aps,
 pra,
]{revtex4-1}
\usepackage{graphicx}% Include figure files
\usepackage{dcolumn}% Align table columns on decimal point
\usepackage{bm}% bold math
\graphicspath{{./figures/}}
\usepackage{amsmath}
\usepackage{color}
\usepackage{physics}
\usepackage{docmute}
\newcommand*{\refequ}[1]{Eq.~(\ref{#1})}
\newcommand*{\up}[1]{^\text{#1}}
\newcommand*{\down}[1]{_\text{#1}}

\begin{document}

%-----------title, authors, affiliations, mailaddres etc.-----------%
\preprint{APS/123-QED}

\title{Implementation of gauge-invariant time-dependent configuration interaction singles method for three-dimensional atoms}% Force line breaks with \\

%author info
\author{Takuma Teramura$^1$} \email{teramura@atto.t.u-tokyo.ac.jp}
\author{Takeshi Sato$^{1, 2, 3}$} \email{sato@atto.t.u-tokyo.ac.jp}
\author{Kenichi L. Ishikawa$^{1, 2, 3}$} \email{ishiken@n.t.u-tokyo.ac.jp}
\affiliation{%
 $^1$Department of Nuclear Engineering and Management, Graduate School of Engineering, The University of Tokyo, 7-3-1 Hongo, Bunkyo-ku, Tokyo 113-8656, Japan
}%
\affiliation{
 $^{2}$Photon Science Center, Graduate School of Engineering, The University of Tokyo, 7-3-1 Hongo, Bunkyo-ku, Tokyo 113-8656, Japan
}%
\affiliation{
 $^{3}$Research Institute for Photon Science and Laser Technology, The University of Tokyo, 7-3-1 Hongo, Bunkyo-ku, Tokyo 113-0033, Japan
}%
%author info

\date{\today}% It is always \today, today,
\begin{abstract}
We present a numerical implementation of the gauge-invariant time-dependent configuration interaction singles (TDCIS) method [Appl.~Sci.~\textbf{8},~433~(2018)] %\cite{Sato_Gauge_2018}
for three-dimensional atoms.
In our implementation, orbital-like quantity called channel orbital [Phys.~Rev.~A \textbf{74},~043420~(2006)] is propagated instead of configuration-interaction (CI) coefficients, which removes a computational bottleneck of explicitly calculating and storing numerous virtual orbitals.
%,  demanding task in simulating electron dynamics induced by intense laser fields in real-space grids.
Furthermore, besides its physical consistency, the gauge-invariant formulation allows to take advantages of the velocity gauge treatment of the laser-electron interaction over the length gauge one in the simulation of high-field phenomena.
We apply the present implementation to high-harmonic generation from helium and neon atoms, and numerically confirms the gauge invariance and demonstrates the effectiveness of the {\it rotated} velocity gauge treatment.
\end{abstract}

\maketitle

\section{\label{sec:level1}Introduction}
Recent laser technology capable of generating strong laser pulses with an intensity $\gtrsim 10^{14}\,{\rm W/cm}^2$ has enabled us to explore electron dynamics in nonperturbative regime, e.g., high-harmonic generation (HHG), above threshold ionization, nonsequential double ionization, and attosecond pulse generation \cite{RevModPhys.81.163, Chem.Rev.6b00453, Annu.Rev.Phys.Chem.2013}.
While laser-driven electron dynamics is rigorously described by the
time-dependent Schr\"odinger equation (TDSE), its direct numerical solution is practically unfeasible for systems with more than two electrons.
For theoretical investigation of multielectron dynamics in intense laser field,
various tractable \textit{ab-initio} methods have been developed,
e.g., time-dependent multiconfiguration self-consistent field (TD-MCSCF) methods \cite{Zanghellini2003, Kato2004,PhysRevA.91.012509,PhysRevA.89.063416, Caillat_MCTDHF, PhysRevA.87.062511,Miyagi_2013,Sato_CAS_2013,Miyagi_2014,Haxton_2015, Sato_ORMAS_2015,Sato_CAS2_2016, Anzaki_MCSCF_2017},
time-dependent coupled cluster method \cite{Kvaal_2012,Sato_OCC_2018},
time-dependent $R-$matrix approach \cite{PhysRevA.79.053411, PhysRevLett.101.253001,Burke_1997,Moore2011, Clarke2018},
and time-dependent reduced two body density matrix approach\cite{PhysRevA.91.023412, PhysRevA.95.033414}.

Among them, the time-dependent configuration interaction singles
(TDCIS) method is one of the promissing methods \cite{Rohringer_TDCIS_2006,PhysRevLett.96.223902, Karam_CISXe_2017, PhysRevLett.106.053003, Greenman_TDCIS_2010, PhysRevA.95.062107, Pabst_JPB_2014, PhysRevA.85.023414,PhysRevA.86.063411,PhysRevLett.111.233005,PhysRevA.89.043415}.
This method has been successfully applied to various electron dynamics such as giant enhancement in HHG in Xe \cite{PhysRevLett.111.233005} and decoherence in attosecond photoionization \cite{PhysRevLett.96.223902}.
In the TDCIS method, the total electronic wavefunction is approximated by a superposition of time-independent Slater determinants
\begin{equation}
\label{eq:sec1:cis}
\ket{\Psi(t)} = \ket{\Phi_0}C_0(t) + \sum^\text{occ}_i\sum^\text{vir}_a
\ket{\Phi^a_i} C^a_i(t),
\end{equation}
where $\ket{\Phi_0}$ is the Hartree-Fock (HF) ground state and $\ket{\Phi^a_i}$ is a singly-excited configuration replacing an occupied orbital $\phi_i$ with a virtual orbital $\phi_a$ unoccupied in the ground state.
The orbital functions are fixed and propagation of configuration-interaction (CI) coefficients ($C_0$ and$\qty{C^a_i}$) describes the system dynamics.
Although applications of the TDCIS method are limited to the single excitation or ionization due to the truncation of CI space, its low computational cost and ease of analysis are attractive.

The conventional TDCIS method with CI coefficients has two major issues;
the explicit calculation and storage of virtual orbitals $\qty{\phi_a}$ and a violation of gauge invariance.
Virtual orbitals $\qty{\phi_a}$ should include both bound and continuum orbitals,
whose number is infinite in principle.
In a practical simulation with real-space grids, one has to prepare virtual orbitals in advance by numerically obtaining all eigenstates of the discretized HF equation.
The number of the virtual orbitals increases with the number of the grid points. Thus, the calculation and storage of virtual orbitals become unacceptably demanding for systems with a large number of grid points like molecules.
To solve this problem, Rohringer \textit{et.al.} have proposed an alternative but equivalent formulation of the TDCIS method in which time-dependent orbital-like quantity called channel orbital is propagated instead of CI coefficient \cite{Rohringer_TDCIS_2006}.
The channel orbital is defined by using $\qty{\phi_a}$ and $C^a_i$ as
\begin{equation}
    \label{eq:sec1:chorb}
    \ket{\chi_i(t)} \equiv \sum^\text{vir}_a \ket{\phi_a}C^a_i(t).
\end{equation}
The equations of motions (EOMs) of CI coefficient are converted to those of $C_0$ and $\qty{\chi_i}$.
This reformulation removes computational bottleneck of handling numerous virtual orbitals, while in princple including all the virtual orbitals within a grid space, and has enhanced utility of the TDCIS method.
% \\ \red{Application(chi)}
However, to the best of our knowledge, the applications of the channel orbital-based TDCIS method have been limited to one-dimensional (1D) model in Refs. \cite{Rohringer_TDCIS_2006,PhysRevA.95.023409, PhysRevA.93.033413} and nobles gas atoms with Hartree-Slater potential in Ref. \cite{PhysRevLett.96.223902}.
% \\ \red{Gauge invariance}

The TDCIS method, either CI coefficient-based or channel-orbital-based, suffers from a violation of gauge invariance,
as a general consequence of the truncation of CI space.
Although it is known that the velocity gauge (VG) offers efficient simulations of high-field phenomena,
it was impossible to enjoy this advantage within the TDCIS method.
To overcome this difficulty, we have recently reported a gauge-invariant reformulation of the TDCIS method \cite{Sato_Gauge_2018}.
In our reformulation, a \textit{rotated} velocity gauge (rVG) transformed from the length gauge (LG) by a unitary operator has been introduced.
This unitary transformation ensures the gauge invariance between the LG and rVG,
and Ref.~\cite{Sato_Gauge_2018} numerically confirmed the equivalence of these gauges for a model 1D Hamiltonian.
%However, the actual computational advantage of the proposed rVG has not yet been assessed for real three-dimensional systems. 
% \\ \red{This work}

In this paper, we report a three dimensional numerical implementation of the gauge-invariant TDCIS method for atoms subject to a linearly polarized laser pulse.
We employ a spherical harmonics expansion of orbital functions with the radial coordinate discretized by a finite-element discrete variable representation (FEDVR)\cite{PhysRevA.62.032706, McCurdy_2004, PhysRevE.73.036708,Schneider2011}.
We apply the present implementation to HHG from helium and neon atoms 
and asses the advantage of the rVG over the LG and VG.
% \\ \red{organization}

This paper proceeds as follows. In Sec.~\ref{sec:level2}, we briefly review the TDCIS methods.
%and describe how to evaluate the expectation values of observables and the ionization probability.
%Sec.~\ref{sec:hamil} defines the system Hamiltonian and summarizes gauge transformation within the TDSE,
%Sec.~\ref{sec:ci} describes the conventional TDCIS method with CI coefficients,
%Sec.~\ref{sec:ch-orb} describes conversion from the TDCIS with CI coefficients to the one with channel orbitals,
%, and the rVG treatment is introduced in Sec.~\ref{sec:rvg}.
The numerical implementation of the gauge-invariant TDCIS method to three-dimensional atoms is given in Sec.~\ref{sec:level3}. 
We describe numerical applications in Sec.~\ref{sec:level4} and conclude this work in Sec.~\ref{sec:level5}.
%In \ref{sec:He}, we compare the LG, VG, and rVG with the TDSE to reveal the gauge invariance of the rVG and LG and the breakdown of the VG.
%In \ref{sec:Ne}, we perform convergence study in neon and show the computational advantage of the rVG treatment over the LG one.
We use Hartree atomic units (a.u.) throughout the paper unless otherwise stated.

\section{\label{sec:level2}Theory}

\subsection{\label{sec:hamil}The System Hamiltonian and Gauge transformation}
 % \red{Hamiltonian}

% First, we define the system Hamiltonian and summarize gauge transformation which connects the LG and VG within the exact TDSE.
% Here, 
We consider an atom with $N$ electrons with a nucleus located at the origin.
The time evolution of the $N$-electron wavefunction $\ket{\Psi(t)}$ is governed by the TDSE,
\begin{equation}
 \label{eq:sec2:tdse}
 i\pdv{}{t}\ket{\Psi(t)} = \hat H(t)\ket{\Psi(t)},
\end{equation}
where $\hat H(t)$ is the time-dependent non-relativistic Hamiltonian
\begin{equation}
 \label{eq:sec2:hamil}
\hat H(t) = \hat H_0 + \hat H\down{ext}(t),
\end{equation}
decomposed into the field-free part,
\begin{equation}
 \hat H_0 = \sum^N_{i = 1}\hat h_0(\vb*{r}_i, \vb*{p}_i) +
  \sum^N_{i > j}\frac{1}{\abs{\vb*{r}_i - \vb*{r}_j}}
\end{equation}
and the laser-electron interaction part
\begin{equation}
 \hat H\down{ext}(t) = \sum^N_{i = 1}\hat h\down{ext}(\vb*{r}_i, \vb*{p}_i, t),
\end{equation}
In these expressions, $\vb*{r}_i$ and $ \vb*{p}_i = -i\nabla_i$ are the position and the canonical momentum of the electron $i$, respectively.
$\hat h_0$ is given by, 
\begin{equation}
 \label{eq:sec2:h0}
 \hat h_0(\vb*{r}, \vb*{p}) = \frac{\vb*{p}^2}{2} -
  \frac{Z}{\abs{\vb*{r}}},
\end{equation}
where $Z$ is the atomic number.
Within the electric dipole approximation, $\hat h\down{ext}$ for the LG and VG are given by
\begin{subequations}
  \label{eq:sec2:Lhamil}
 \begin{align}
  \label{eq:sec2:LG}
  \hat h^\text{LG}_\text{ext}(\vb*{r}, \vb*{p}, t) &= \vb*{E}(t)\cdot\vb*{r} \\
  \label{eq:sec2:VG}
  \hat h^\text{VG}_\text{ext}(\vb*{r}, \vb*{p}, t) &= \vb*{A}(t)\cdot\vb*{p},
  \end{align}
\end{subequations}
where $\vb*{E}(t)$ and $\vb*{A}(t) = -\int^t_{-\infty} dt^\prime\vb*{E}(t^\prime)$ are the electric field and the vector potential of the external laser field, respectively.

% \red{Gauge transformation}
The two gauges are physically equivalent, and any physical observable takes the same value, independent of the choice of the gauge.
% The TDSE for the LG is written as
% \begin{equation}
%   i\frac{\partial}{\partial t}\ket{\Psi\up{LG}(t)}=\hat H\up{LG}(t)\ket{\Psi\up{LG}(t)},
% \end{equation}
The LG wavefunction $\ket{\Psi\up{LG}}$ and VG wavefunction $\ket{\Psi\up{VG}(t)}$ 
% are the wavefunction and the Hamiltonian for the LG, respectively.
% This LG TDSE is connected to the VG TDSE
% $i\partial_t\ket{\Psi\up{VG}(t)} = \hat H\up{VG}(t)\ket{\Psi\up{VG}(t)}$,
% and the VG wavefunction $\ket{\Psi\up{VG}}$ is 
are mutually transformed by a gauge transformation as,
\begin{align}
  \label{eq:sec2:gauge}
&\ket{\Psi\up{VG}(t)} = \hat U(t)\ket{\Psi\up{LG}(t)} \\
&\hat U(t) \equiv \exp\qty[-i\sum^N_{i = 1}\qty(\vb*{A}(t)\cdot \vb*{r}_i
 - \frac{1}{2}\int^t_{-\infty}dt^\prime \abs{\vb*{A}(t^\prime)}^2)]
\end{align}
%%%%%%%%%%%%%%%%%%%%%%%%%%%%%%%%%%%%%%%%%%%%%%%%%%%%%%%%%%%%
\subsection{\label{sec:ci}The CI coefficient-based TDCIS method in the length gauge}
In the conventional TDCIS method based on CI coefficients, orbitals satisfy the canonical, restricted HF equation
\begin{align}\label{eq:sec2:HF}
 \hat f \ket{\phi_p} &\equiv \hat h_0\ket{\phi_p} + \sum^\text{occ}_i(2\hat W^{\phi_i}_{\phi_i}\ket{\phi_p} - \hat W^{\phi_i}_{\phi_p}\ket{\phi_i})\nonumber \\
   &= \epsilon_p \ket{\phi_p},
\end{align}
where $\hat f$ is the Fock operator and $\hat W^{\phi}_{\phi^\prime}$ is the potential due to the product of two given orbitals $\phi$ and $\phi^\prime$, defined in the real space as
\begin{equation}
    \label{eq:sec2:W}
 \hat W^{\phi}_{\phi^\prime}(\vb*{r}_1) 
 \equiv \int d\vb*{r}_2 \frac{\phi^*(\vb*{r}_2)\phi^\prime(\vb*{r}_2)}{\abs{\vb*{r}_1 - \vb*{r}_2}}.
\end{equation}
$\epsilon_p$ is the orbital energy of orbital $\phi_p$.
In the TDCIS wavefunction in \refequ{eq:sec1:cis},
$\ket{\Phi_0}$ is the HF ground state formed with the occupied orbitals as
\begin{equation}
 \ket{\Phi_0} = \prod\up{occ}_i \hat c^\dagger_{i\uparrow}\hat c^\dagger_{i\downarrow} \ket{},
\end{equation}
where $\hat c^\dagger_{p\sigma}$ and $\hat c_{p\sigma}$ are the creation and annihilation operators, respectively, of spin-orbital $\phi_p\otimes\sigma$,
and $\ket{}$ is the vacuum. $\sigma \in\qty{\uparrow,\downarrow}$ denotes the spin function.
 $\ket{\Phi^a_i}$ is a singly-excited configuration replacing an occupied orbital $\phi_i$ with a virtual orbital $\phi_a$
\begin{equation}
 \ket{\Phi^a_i} = \frac{1}{\sqrt{2}}(\hat
  c^\dagger_{a\uparrow}\hat c_{i\uparrow} + \hat c^\dagger_{a\downarrow}\hat c_{i\downarrow}) \ket{\Phi_0}.
\end{equation}

The EOMs of CI coefficients is derived through the Dirac-Frenkel time-dependent variational principle \cite{Kramer:107648},
requiring Lagrangian $L(t)$
\begin{equation}
 L(t) = \bra{\Psi}\hat H(t) - i\pdv{}{t} - E_0 \ket{\Psi}
\end{equation}
to be stationary with respect to the variation of $C^*_0$ and $\qty{C^{a*}_i}$.
% Thus,
% \begin{equation}
%   \label{eq:sec2:tdvp}
%  \pdv{L}{C^*_0} = \pdv{L}{C^{a*}_i} = 0.
% \end{equation}
$E_0=\bra{\Phi_0}\hat H_0 \ket{\Phi_0}$ denotes the HF energy.
% given as
% \begin{align}
%  E_0 &= \bra{\Phi_0}\hat H_0 \ket{\Phi_0} \nonumber \\
%  &= 2\sum^\text{occ}_i \epsilon_i
%  -\sum\up{occ}_{ij} (2\bra{\phi_i}\hat W^{\phi_j}_{\phi_j}\ket{\phi_i}
%   - \bra{\phi_i}\hat W^{\phi_j}_{\phi_i}\ket{\phi_j}).
% \end{align}
This constant shift, introduced for the simple notation of the EOMs, does not affect physical results.
% \red{TDVP}
In the LG case, in which the wavefuntion is written as,
 \begin{equation}
   \label{eq:sec2:TDCISLG}
   \ket{\Psi\up{LG}(t)} = \ket{\Phi_0}C_0 + \sum\up{occ}_i\sum\up{vir}_a \ket{\Phi^a_i}C^a_i.
 \end{equation}
% The stationary condition (Eq. \ref{eq:sec2:tdvp}) is equivalent to
% \begin{subequations}
% \begin{align}
%  \bra{\Phi_0}\qty(\hat H\up{LG}(t) - i\partial_t)\qty( \ket{\Phi_0}C_0 + \sum\up{occ}_j\sum\up{vir}_b \ket{\Phi^b_j}C^b_j) &= 0 \\
%   \bra{\Phi^a_i}\qty(\hat H\up{LG}(t) - i\partial_t)\qty( \ket{\Phi_0}C_0 + \sum\up{occ}_j\sum\up{vir}_b \ket{\Phi^b_j}C^b_j) &= 0.
% \end{align}
% \end{subequations}
% By using Slater-condon rules and Brillouin theorem \cite{Szabo_ModQuantChem},
the EOMs of CI coefficients are obtained as
\begin{subequations}
  \label{eq:sec2:ciLG}
 \begin{align}
  i\dot C_0 &=
   \sqrt{2}\vb*{E}\cdot\sum^\text{occ}_{j}\sum^\text{vir}_{b}
   \bra{\phi_j} \vb*{r}\ket{\phi_b}C^b_j \\
  i\dot C^a_i 
  &= \bra{\phi_a}\{
  \sum^\text{occ}_j
  \sum^\text{vir}_b
  \hat{F}_{ij}
  \ket{\phi_b}C^b_j
  +\sum^\text{vir}_b
  \vb*{E}\cdot\vb*{r}\ket{\phi_b}C^b_i 
  \nonumber \\
  &+ \sqrt{2} \vb*{E}\cdot\vb*{r}\ket{\phi_i}C_0\} 
  - \vb*{E}\cdot\sum^\text{occ}_j \bra{\phi_j} \vb*{r}\ket{\phi_i}C^a_j,
 \end{align}
\end{subequations}
where
\begin{equation}
  \label{eq:sec2:Fij}
 \hat F_{ij}\ket{\phi_b} = \delta^i_j(\hat f - \epsilon_i) \ket{\phi_b} + 
2\hat W^{\phi_j}_{\phi_b}\ket{\phi_i} - \hat
  W^{\phi_j}_{\phi_i}\ket{\phi_b}.
\end{equation}
%%%%%%%%%%%%%%%%%%%%%%%%%%%%%%%%%%%%%%%%%%%%%%%%%%%%%%%%%%%%
\subsection{The channel orbital-based \label{sec:ch-orb}TDCIS method in the length gauge}
The EOMs of CI coefficients [Eq. (\ref{eq:sec2:ciLG})] can be rewritten, by substituting channel orbital Eq. (\ref{eq:sec1:chorb}) into Eq. (\ref{eq:sec2:ciLG}), as,
\begin{subequations}
\label{eq:sec2:chorbLG}
\begin{align}
 i\dot C_0 &= \sqrt{2}\vb*{E}\cdot\sum^\text{occ}_j\bra{\phi_j}
 \vb*{r}\ket{\chi_j} \\
 i\ket{\dot \chi_i} &=
  \hat P \qty{ (\hat F + \vb*{E}\cdot\vb*{r})\ket{\chi_i} + \sqrt{2}
  \vb*{E}\cdot\vb*{r} \ket{\phi_i}C_0} \nonumber \\
&- \vb*{E}\cdot\sum^\text{occ}_j
  \ket{\chi_j}\bra{\phi_j} \vb*{r}\ket{\phi_i},
\end{align}
\end{subequations}
where
\begin{equation}
  \label{eq:sec2:Fi}
 \hat F\ket{\chi_i} = (\hat f - \epsilon_i) \ket{\chi_i} + \sum^\text{occ}_j
(2\hat W^{\phi_j}_{\chi_j}\ket{\phi_i} - \hat
  W^{\phi_j}_{\phi_i}\ket{\chi_j}),
\end{equation}
and $\hat P$ is the projection operator onto the space spanned by virtual orbitals
\begin{equation}
  \label{eq:sec2:P}
 \hat P = \sum^\text{vir}_a\ket{\phi_a}\bra{\phi_a} = \hat 1 - \sum^\text{occ}_j\ket{\phi_j}\bra{\phi_j},
\end{equation}
with $\hat 1$ being the identity operator.
%%%%%%%%%%%%%%%%%%%%%%%%%%%%%%%%%%%%%%%%%%%%%%%%%%%%%%%%%%%%
\subsection{\label{sec:rvg}Velocity gauge and rotated velocity gauge}
One can, in principle, derive the EOMs for the VG case in the same way as for the LG.
Let us write the total wavefunction and channel orbital as,
\begin{align}
  \ket{\Psi\up{VG}(t)} &= \ket{\Phi_0}D_0 + \sum\up{occ}_i\sum\up{vir}_a \ket{\Phi^a_i}D^a_i \\
  \ket{\eta(t)} &= \sum\up{vir}_a \ket{\phi_a}D^a_i.
\end{align}
$\ket{\Phi_0}$ and $\ket{\Phi^a_i}$ are the same configurations as those used in the LG case.
Their EOMs are obtained through the same procedures as in the LG case as,
\begin{subequations}
\label{eq:sec2:chorbVG}
\begin{align}
 i\dot D_0 &= \sqrt{2}\vb*{A}\cdot\sum^\text{occ}_j\bra{\phi_j}
   \vb*{p}\ket{\eta_j} \\
 i\ket{\dot \eta_i} &=
  \hat P \qty{ (\hat F +  \vb*{A}\cdot \vb*{p})\ket{\chi_i} + \sqrt{2}
   \vb*{A}\cdot \vb*{p} \ket{\phi_i}C_0} \nonumber \\
&- \vb*{A}\cdot\sum^\text{occ}_j
  \ket{\eta_j}\bra{\phi_j} \vb*{p}\ket{\phi_i}.
\end{align}
\end{subequations}

It is known that TDCIS, which uses time-independent orbitals, is not gauge-invariant \cite{Ishikawa2015, Sato_Gauge_2018, PUILS}.
% We note that all orbitals are fixed and the VG wavefunction is not connected with the LG one by gauge transformation operator $\hat U(t)$.
% Truncation of CI space results in a violation of gauge invariance.
% Except full-CI expansion in which all electric configurations are considered, all TDCI methods in truncated CI space with fixed orbitals are gauge dependent.
Instead of the conventional VG as described above, we have recently proposed the rVG \cite{Sato_Gauge_2018}, where
we define the rVG wavefunction by the gauge transformation from the LG wavefunction as, 
 \begin{equation}
   \ket{\Psi\up{rVG}(t)} = \hat U(t)\ket{\Psi\up{LG}(t)}
 \end{equation}
The rVG orbitals are related to the LG ones by,
\begin{align}
\label{eq:sec2:rvgphi}
  \ket{\phi^\prime_p(t)} &= \hat u(t)\ket{\phi_p} \\
\label{eq:sec2:rvgchi} 
  \ket{\chi^\prime_i(t)} &= \hat u(t)\ket{\chi_i} = \sum\up{vir}_a\ket{\phi^\prime_a}C^a_i,
\end{align}
where
\begin{equation}
 \hat u(t) = \exp{-i\qty(\vb*{A}(t)\cdot\vb*{r} -
  \frac{1}{2}\int^t_{-\infty}dt^\prime \abs{\vb*{A}(t^\prime)}^2)}.
\end{equation}
They satisfy the following EOMs \cite{Sato_Gauge_2018}:
\begin{subequations}
\label{eq:sec2:rvg}
\begin{align}
 i\dot C_0 &=
  \sqrt{2}\vb*{E}\cdot\sum^\text{occ}_j\bra{\phi^\prime_j}\vb*{r}\ket{\chi^\prime_j} \\
 i\ket{\dot \chi^\prime_i} &=
  \hat P^\prime \qty{ (\hat F^\prime + \vb*{A}\cdot\vb*{p})\ket{\chi^\prime_i} + \sqrt{2}\vb*{E}\cdot\vb*{r}\ket{\phi^\prime_i}C_0 } \nonumber \\
&-\vb*{E}\cdot\sum^\text{occ}_j
  \qty(\ket{\chi^\prime_j}\bra{\phi^\prime_j}\vb*{r}\ket{\phi^\prime_i}
  +  \ket{\phi^\prime_j}\bra{\phi^\prime_j}\vb*{r}\ket{\chi^\prime_i}),
\nonumber \\
\end{align}
\end{subequations}
where $\hat P^\prime$ and $\hat F^\prime_i$ are given by Eqs.~ (\ref{eq:sec2:P}) and (\ref{eq:sec2:Fi}), respectively, with $\qty{\phi_j}$ replaced with $\qty{\phi^\prime_j}$.
Although Eq.~(\ref{eq:sec2:rvg}) contains the dipole operator $\vb*{E}\cdot \vb*{r}$, this does not prevent enjoying the advantages of the VG treatment,
since it acts only on the localized occupied orbitals $\{\phi^\prime_i\}$.
% The equivalence of Lagrangian of the rVG and LG is confirmed by seeing 
% \begin{align}
%   L\up{LG} (t)
%   &= \bra{\Psi\up{rVG}}\hat U
%   \qty(\hat H_0 + \hat H\up{LG}\down{ext} - E_0 -i\partial_t )\hat U^\dagger
%   \ket{\Psi\up{rVG}} \nonumber \\
%   &= \bra{\Psi\up{rVG}}
%   \qty(\hat H_0 + \hat H\up{VG}\down{ext} - E_0 -i\partial_t )
%   \ket{\Psi\up{rVG}} \nonumber \\
%  &= L\up{rVG}(t).
% \end{align}
%%%%%%%%%%%%%%%%%%%%%%%%%%%%%%%%%%%%%%%%%%%%%%%%%%%%%%%%%%%%

%%%%%%%%%%%%%%%%%%%%%%%%%%%%%%%%%%%%%%%%%%%%%%%%%%%%%%%%%%%%
%%%%%%%%3.Implementation%%%%%%%%
\section{\label{sec:level3}Implementation to three-dimensional atoms}
\subsection{\label{sec:FEDVR}Spherical-FEDVR basis}
  %We have successfully implemented the gauge-invariant TDCIS method for multielectron dynamics driven by linearly polarized laser field.
  The present implementation is based on our TD-MCSCF code \cite{Sato_CAS2_2016}, which uses spherical-FEDVR basis functions
  \begin{equation}
    \label{eq:sec3:sphFEDVR}
      \psi_{klm}(\vb*{r}) = \frac{1}{r}\alpha_k(r)Y_{lm}(\Omega),
  \end{equation}
 where $r$ and $\Omega$ are the radial and angular coordinate of $\vb*{r}$, respectively,
 $Y_{lm}$ are spherical harmonics, and
 $\alpha_k$ are radial FEDVR basis functions \cite{PhysRevA.62.032706, McCurdy_2004}.
 The radial coordinate of the simulation box $[0, R\down{max}]$ is divided into $K\down{FE}$ finite elements.
 Each finite element supports $K\down{DVR}$ local DVR functions, and neighboring elements are connected by a bridge function.
In total, there are $K\down{rad} = K\down{FE}K\down{DVR} - (K\down{FE} - 1)$ radial grid points $\{r_k\}$, on which 
$
    \alpha_k(r_{k^\prime}) = \delta_{kk^\prime}/\sqrt{w_k},
$
with $\{w_k\}$ being the integration weights.

We expand the channel orbital $\chi_i$ in the spherical-FEDVR basis as
\begin{equation}
  \chi_i(\vb*{r};t) = \sum^{K\down{rad} - 1}_{k = 2}\sum^{L\down{max}}_{l = 0}\psi_{klm_i}(\vb*{r})g^{kl}_i(t),
\end{equation}
where $L\down{max}$ is the maximum angular momentum included. The FEDVR basis functions corresponding to $r_1 = 0$ and $r_{K\down{rad}} = R\down{max}$ are removed to enforce the vanishing boundary condition for $r\chi_i$ at both ends of the simulation box.
%In our simulations, $g^{kl}_i$ is the working variable and the EOMs in the basis of the spherical-FEDVR is solved.

The electrostatic potentials for electron-electron interaction, $\hat W^{\phi_j}_{\phi_i}(\vb*{r})$ and $W^{\phi_j}_{\chi_i}(\vb*{r},t)$ required for the EOM of channel orbitals, are computed by solving Poisson's equation,
\begin{subequations}
\label{eq:sec3:poi}
\begin{align}
    \label{eq:sec3:poi1}
    \nabla^2 \hat W^{\phi_j}_{\phi_i}(\vb*{r}) &= -4\pi\phi_j^*(\vb*{r})\phi_i(\vb*{r}), \\
    \label{eq:sec3:poi2}
    \nabla^2 \hat W^{\phi_j}_{\chi_i}(\vb*{r},t) &= -4\pi\phi_j^*(\vb*{r})\chi_i(\vb*{r},t),
\end{align}
\end{subequations}
using the method described in Ref.~\cite{Sato_CAS2_2016}.
It should be noted that $\hat W^{\phi_j}_{\phi_i}(\vb*{r})$ is time independent, and Eq.~(\ref{eq:sec3:poi1}) needs to be solved only once before the simulation.
On the other hand, $\hat W^{\phi_j}_{\chi_i}(\vb*{r},t)$ is time dependent, and should be computed at every time step. However, since 
its source $\phi^*_j(\vb*{r})\chi_i(\vb*{r},t)$ [See Eq.~(\ref{eq:sec3:poi2}).] and operand $\{\phi_j(\vb*{r})\}$ [See Eq.~(\ref{eq:sec2:Fi})] are both localized around the atom due to the locality of occupied orbitals, Eq.~(\ref{eq:sec3:poi2}) can be solved with less computational cost than the similar equation appearing, e.g, in time-dependent Hartree-Fock and TD-MCSCF method \cite{Sato_CAS2_2016}.

\subsection{\label{sec:td}Time-propagation with exponential time differencing fourth-order Runge-Kutta scheme}
For an efficient propagation of the EOM of channel-orbital based TDCIS, we use the exponential time differencing fourth-order Runge-Kutta scheme (ETDRK4) by Krogstad \cite{Krogstad2005, hochbruck_ostermann_2010,Kidd2017}. To this end, we arrange $C_0$ and $\qty{\chi_i}$ into a unified vector $\vb*{\chi} = (C_0, \chi)^T$
and rewrite the EOMs of $C_0$ and $\qty{\chi_i}$ by a matrix form 
\begin{equation}
    i\pdv{}{t}\vb*{\chi} = \vb*{h}\vb*{\chi} +\vb*{W}[\vb*{\chi}, t],
\end{equation}
where $\vb*{h}$ is a chosen stiff part of the right-hand side of the EOM (See below), and $\vb*{W}[\vb*{\chi}, t]$ is a nonstiff remainder. We choose the stiff part $\vb*{h}$ to be either (i) the field-free one-electron Hamiltonian $\hat{h}\down{0}$ or (ii) the totality of the one-electron Hamiltonian $\hat{h}\down{0}+\hat{h}\down{ext}(t)$. For the first case (i) with time-independent $\vb*{h}$, 
%In our EOMs, there are two stiff parts, $\hat h_0$ [Eq.~(\ref{eq:sec2:h0})] and $\hat h\down{ext}(t)$ [Eq.~(\ref{eq:sec2:Lhamil})].
%The former is to be put into $\vb*{h}$.
%However, one should not directly put $\hat h\down{ext}$ into $\vb*{h}$ because of its time-dependence.
%To treat this time-dependent linear term within ETDRK4,
%$\hat h\down{ext}$ is divided into time-independent part and time-dependent part \cite{hochbruck_ostermann_2010},
%and time-independent part is put into $\vb*{h}$.
%We put time-dependent part of $\hat h\down{ext}$ and
%the rest of the EOMs are absorbed in $\vb*{W}$.
the time propagation from $\vb*{\chi}_n = \vb*{\chi}(t_n)$ to $\vb*{\chi}_{n + 1} = \vb*{\chi}(t_n + \Delta t)$ is given by 
\begin{align}
    \vb*{\chi}_{n + 1} &= \varphi_0\qty(-i\vb*{h}\Delta t)\vb*{\chi}_n
    -i\Delta t \left[
    f_0(-i\vb*{h}\Delta t)\vb*{W}_n \right.\nonumber \\
    &\left.+ f_1(-i\vb*{h}\Delta t)(\vb*{W}_a + \vb*{W}_b) 
    + f_2(-i\vb*{h}\Delta t)\vb*{W}_c
    \right],
\end{align}
where $f_1$, $f_2$, and $f_3$ are defined as
\begin{subequations}
\begin{align}
f_0(z) &= \varphi_1(z) - 3\varphi_2(z) + 4\varphi_3(z) \\
f_1(z) &= 2\varphi_2(z) - 4 \varphi_3(z) \\
f_2(z) &= -\varphi_2(z) + 4\varphi_3(z) ,
\end{align}
\end{subequations}
where $z=-i\vb*{h}\Delta t$, $\varphi_0(z)=e^z$, and
\begin{eqnarray}\label{eq:phirec}
 \varphi_{k + 1}(z) = \frac{1}{z}\qty(\varphi_l(z) - \frac{1}{k!}) \hspace{1em} (k=0,1,2,\cdots).
\end{eqnarray}
$\vb*{W}_n$, $\vb*{W}_a$, $\vb*{W}_b$, and $\vb*{W}_c$ are given by
\begin{subequations}    
\begin{align}
    \vb*{W}_n &= \vb*{W}[\vb*{\chi}_n, t_n] \\
    \vb*{W}_a &= \vb*{W}[\vb*{a}_n, t_{n} + \Delta t/2] \\
    \vb*{W}_b &= \vb*{W}[\vb*{b}_n, t_{n} + \Delta t/2] \\
    \vb*{W}_c &= \vb*{W}[\vb*{c}_n, t_{n + 1}],
\end{align}
\end{subequations}
where $\vb*{a}_n$, $\vb*{b}_n$, and $\vb*{c}_n$ are intermediate vectors given as 
\begin{subequations}    
\begin{align}
\vb*{a}_n &= \varphi_0 \qty(z/2)\vb*{\chi}_n 
-i \Delta t\varphi_1 \qty(z/2)\vb*{W}_n/2 \\
\vb*{b}_n &= \varphi_0 \qty(z/2)\vb*{\chi}_n 
-i \Delta t\varphi_1 \qty(z/2)\vb*{W}_n/2 \nonumber \\
 &-i\Delta t\varphi_2\qty(z/2)(\vb*{W}_a - \vb*{W}_n) \\
\vb*{c}_n &= \varphi_0\qty(z)\vb*{\chi}_n -i\Delta t\varphi_1\qty(z)\vb*{W}_n -2i\Delta t \varphi_2(z)(\vb*{W}_b - \vb*{W}_n)
\end{align}
\end{subequations}    
% Cox and Mathews
% \begin{align}
% \vb*{a}_n &= e^{-i\frac{\vb*{h}\Delta t}{2}}\vb*{\chi}_n 
% -i \vb*{h}^{-1}(e^{-i\vb*{h}\Delta t} - \vb*{I})\vb*{W}[\vb*{\chi}_n, t_n] \\
% \vb*{b}_n &= e^{-i\frac{\vb*{h}\Delta t}{2}}\vb*{\chi}_n 
% -i\vb*{h}^{-1}(e^{-i\vb*{h}\Delta t} - \vb*{I})\vb*{W}[\vb*{a}_n, t_{n + \frac{1}{2}}] \\
% \vb*{c}_n &= e^{-i\vb*{h}\Delta t}\vb*{a}_n\nonumber\\
% &-i \vb*{h}^{-1}(e^{-i\vb*{h}\Delta t} - \vb*{I})(2\vb*{W}[\vb*{b}_n, t_{n + \frac{1}{2}}] - \vb*{W}[\vb*{\chi}_n, t_n]).
% \end{align}
The operator exponential $\varphi_0(z)$ and $\varphi_0(z/2)$ in the spherical-FEDVR basis are approximated by the Pad\'e~(3/3) approximation,
and higher-order $\varphi_k$ functions are obtained by successively applying Eq.~(\ref{eq:phirec}).
The denominator of the Pad\'e approximation is factorized and operated by the matrix iteration method \cite{Sato_CAS2_2016}.
We follow Ref.~\cite{hochbruck_ostermann_2010} for the modification required for a time-dependent stiff part $\vb*{h}$.
In the absence of linear part for $C_0$, time-propagation of $C_0$ reduces to well-known fourth-order Runge-Kutta scheme.

\begin{widetext}
\subsection{\label{sec:eval}Expectation value}
% Let us consider the way to compute 
The expectation value of one-body operator $\ev{O} = \bra{\Psi}O\ket{\Psi}$ can be evaluated in the LG case as \cite{Rohringer_TDCIS_2006,Sato_Gauge_2018} 
% and its time derivative $\dot{\ev{O}} = d\ev{O}/dt$ .
%In the LG case, $\ev{O}$ is computed by
\begin{align}
 \label{eq:sec2:o}
 \ev{O} &= \sum\up{occ}_i\qty{2\bra{\phi_i}O\ket{\phi_i}
 + \bra{\chi_i}O\ket{\chi_i}} 
 +2\sqrt{2}\Re\qty[C_0\sum\up{occ}_i\bra{\chi_i}O\ket{\phi_i}] 
 -\sum\up{occ}_{ij}\ip{\chi_i}{\chi_j}\bra{\phi_j}O \ket{\phi_i}. 
\end{align}
The VG expression is obtained by simply replacing $\qty{C_0, \chi_j}$ with $\qty{D_0, \eta_j}$,
and the rVG one by replacing $\qty{\phi_j, \chi_j}$ with $\qty{\phi^\prime_j, \chi^\prime_j}$.
The Ehrenfest theorem $\frac{d}{dt}\ev{O}=-i \bra{\Psi}[O, \hat H]\ket{\Psi}$ does not hold for TDCIS. 
% To compute $\dot{\ev{O}}$ from the exact solution of the TDSE, 
% \begin{subequations}
%  \begin{align}
% \label{eq:sec2:odot}
%   \frac{d\bra{\Psi}O\ket{\Psi}}{dt} &= \bra{\Psi}O\dot{\ket{\Psi} }
%   + \dot{\bra{\Psi}}O \ket{\Psi} \\
% \label{eq:sec2:ehren}
% &= -i \bra{\Psi}[O, \hat H]\ket{\Psi}.
%  \end{align}
% \end{subequations}
% is used.
% The last relation, \refequ{eq:sec2:ehren}, is known as the Ehrenfest theorem.
% However, this theorem does not hold in the TDCIS method.
Instead, we evaluate the time derivative of $\ev{O}$ as \cite{Sato_Gauge_2018},
% Thus, one must explicitly evaluate \refequ{eq:sec2:odot} to calculate $\dot{\ev{O}}$.
% In the LG case, \refequ{eq:sec2:odot} is derived as
\begin{align}
\label{eq:sec2:doto}
 \dot{\ev{O}} \equiv \frac{d\ev{O}}{dt} &=
 2\Re\qty[\sum_i\up{occ}\qty{
 \bra{\chi_i}O\ket{\dot \chi_i} + \sqrt{2}\dot C_0\bra{\chi_i}O\ket{\phi_i} 
+ \sqrt{2}C_0\bra{\dot \chi_i}O\ket{\phi_i}
}] 
\end{align}
in the LG case. $\qty{C_0, \chi_j}$ is to be replaced with $\qty{D_0, \eta_j}$ for VG.
The rVG case needs extra terms \cite{Sato_Gauge_2018}:
\begin{align}
\label{eq:sec2:dotorvg}
\dot {\ev{O}} = 
&2\Re\qty[\sum\up{occ}_i\qty{
\bra{\chi^\prime_i}O\ket{\dot \chi^\prime_i} 
+ \sqrt{2}\dot C_0\bra{\chi^\prime_i}O\ket{\phi^\prime_i} 
+ \sqrt{2}C_0\bra{\dot \chi^\prime_i}O\ket{\phi^\prime_i}
}- \sum\up{occ}_{ij}\ip{\chi^\prime_i}{\dot \chi^\prime_j}\bra{\phi^\prime_j}O\ket{\phi^\prime_i}]\nonumber \\
&
 - \sqrt{2}\Im \qty[
 C_0\sum\up{occ}_i \qty{2\vb*{E}\cdot \bra{\chi^\prime_i}O \vb*{\hat r}\ket{\phi^\prime_i}
 +\abs{\vb*{A}}^2\bra{\chi^\prime_i}O\ket{\phi^\prime_i}}
 ] 
 - i\vb*{E}\cdot\sum\up{occ}_{ij}\qty(2\delta^i_j - \ip{\chi^\prime_i}{\chi^\prime_j})\bra{\phi^\prime_j}[\vb*{\hat r}, O]\ket{\phi^\prime_i}.
\end{align}
Equations (\ref{eq:sec2:o}), (\ref{eq:sec2:doto}), and (\ref{eq:sec2:dotorvg}) are valid not only for atoms but also any multielectron system including molecules.
\end{widetext}

\subsection{Ionization probability}
%Another important property to be extracted from the system wavefunction is a measure of the ionization yield. 
To conveniently analyze how ionization proceeds using the TD-MCSCF wavefunctions with time-varying orbitals,
%in many electron systems,
%In the TDCIS method, only neutral and singly-ionized states are to be populated.
%we have introduced a real-space domain-based classification of different charged states \cite{Sato_CAS_2013}.
we have previously introduced \cite{Sato_CAS_2013} a domain-based $n$-fold ionization probability $P_n$, defined as a probability to find $n$ electrons in the outer region $\abs{\vb*{r}} > R\down{ion}$ and the other $N-n$ electrons in the inner region $\abs{\vb*{r}} < R\down{ion}$ with a given distance $R\down{ion}$ from the origin.
This quantity is gauge-invariant even during the pulse, unlike the population of the (field-free) continuum levels \cite{Ishikawa2015,PUILS}.
% a classification of different charged states based on the electron position in the real space,  and applied to the TD-MCSCF wavefunction
%We regard a condition where all electrons exist near the nuclei as the neutral state.
% The yield of the neural state $P_0$, e.g, is defined as the probability to find all electrons in the inner region $\abs{\vb*{r}} < R\down{ion}$ with a given distance $R\down{ion}$,
% \begin{equation}
%  P_0(t) \equiv \int_< dx_1 \cdots \int_< dx_N \abs{\Psi(x_1,\ldots,x_N;t)}^2,
% \end{equation}
% where $x = \qty{\vb*{r}, \xi}$ is the spatial-spin coordinate and $\int_< dx$ is the integration in the inner region,
% \begin{equation}
% \int_< dx = \int_{\abs{\vb*{r}} < R\down{ion}} d\vb*{r}\int d\xi.
% \end{equation}
To apply this approach to the TDCIS method with channel orbitals, it is reasonable to assume that 
the occupied orbitals $\{\phi_i\}$ are localized inside the inner region, i.e, $\phi_i(\vb*{r}) = 0$ for $\abs{\vb*{r}} > R\down{ion}$. 
Then, the yield of the neutral species, or the survival probability $P_0$ is computed as,
\begin{equation}
 P_0(t) = \abs{C_0(t)}^2 + \sum\up{occ}_i \ip{\chi_i}{\chi_i}_<,
\end{equation}
where $\ip{\chi_i}{\chi_j}_<$ is the overlap integral in the inner region
\begin{equation}
\label{eq:sec2:p0}
\ip{\chi_i}{\chi_j}_< \equiv 
\int_{\abs{\vb*{r}}< R\down{ion}}d\vb*{r}\chi^*_i(\vb*{r};t)\chi_j(\vb*{r};t).
\end{equation}
Noting that the atom described by a TDCIS wavefunction is at most singly ionized, we obtain the single-ionization probability as $P_1(t)=1-P_0(t)$.

% \red{
%  Indeed one can also define ionization yield by projection onto continuum states, but continuum states do not fulfill the requirement of gauge-invariance in the presence of an external laser field. Thus, to compare a degree of ionization with different gauge treatments, we employ the domain-based ionization probability which is gauge-invariant.
% }
%%%%%%%%%%%%%%%%%%%%%%%%%%%%%%%%%%%%%%%%%%%%%%%%%%%%%%%%%%%%
%%%%%%%%4.Numerical example%%%%%%%%
 \section{\label{sec:level4}numerical example}
We present numerical applications of the implementation of the reformulated TDCIS method described in the previous section and assess efficiency of the rVG.
In all simulations reported below, we assume a laser field linearly polarized along the $z$ axis of the following form:
\begin{equation}
  E(t) = \sqrt{I_0}\sin(\omega t)\sin^2\qty(\pi\frac{t}{N\down{opt}T})\,(0 \leq t \leq N\down{opt}T),
\end{equation}
where $I_0$ is the peak intensity,
$\omega$ is the central frequency, $T = 2\pi/\omega$ is the period,
and $N\down{opt}$ is the total number of optical cycles.

%%%%%%%%%%%%%%%%%%%%%%%%%%%%%%%%%%%%%%%%%%%%%%%%%%%%%%%%%%%%x
\subsection{\label{sec:He}Helium}
First, we consider helium atom exposed to a laser pulse with $I_0 = 4.0\times10^{14}\ \text{W/cm}^2$, $\lambda = 400\ \text{nm}$, and $N\down{opt} = 12$.
In this condition, exact numerical solution of the TDSE is available \cite{PhysRevLett.103.063002, PhysRevA.83.053418},
% The TDSE code developed by J. Burgd\"orfer and his collaborators at Vienna University of Technology \cite{PhysRevLett.103.063002, PhysRevA.83.053418} and the present results of the TDSE are provided by them.
% This code was previously used to asses the performance of the TD-CASSCF method \cite{Sato_CAS2_2016}.
%As noted previously, the TDSE obeys the Ehrenfest theorem 
%(Eq. (\ref{eq:sec2:ehren})).
from which the expectation value of dipole velocity and dipole acceleration can be calculated by using the Ehrenfest theorem.
% \begin{subequations}
% \begin{align}
%   \frac{d}{dt}\ev{\hat z} &= \bra{\Psi}\hat p_z\ket{\Psi}\\
%   \frac{d^2}{dt^2}\ev{\hat z} &= -\bra{\Psi}\qty(\pdv{\hat V\down{nuc}}{z} + \pdv{\hat H_\text{ext}}{z})\ket{\Psi},
% \end{align}
% \end{subequations}
% where $\hat V\down{nuc} = -Z/\abs{\vb*{r}}$ is the nuclear potential and $\hat H\down{ext}$ is the laser potential defined in \refequ{eq:sec2:Lhamil}.
For the TDCIS method,
we apply $O = \hat z$ in Eqs.~(\ref{eq:sec2:o}) and (\ref{eq:sec2:doto}) to evaluate the expectation value of dipole moment and velocity, respectively.
%Due to the symmetry of occupied orbital in atoms, 
%$\bra{\phi}\hat z\ket{\phi} = 0$ vanishes some terms in Eq. (\ref{eq:sec2:o}, \ref{eq:sec2:doto}).
Dipole acceleration is computed by numerically differentiating dipole velocity.

\begin{figure}[ht]
    %---------------------------------------------------%
    \begin{minipage}[htp]{1.0\hsize}
      \centering
      \includegraphics[width = \textwidth]{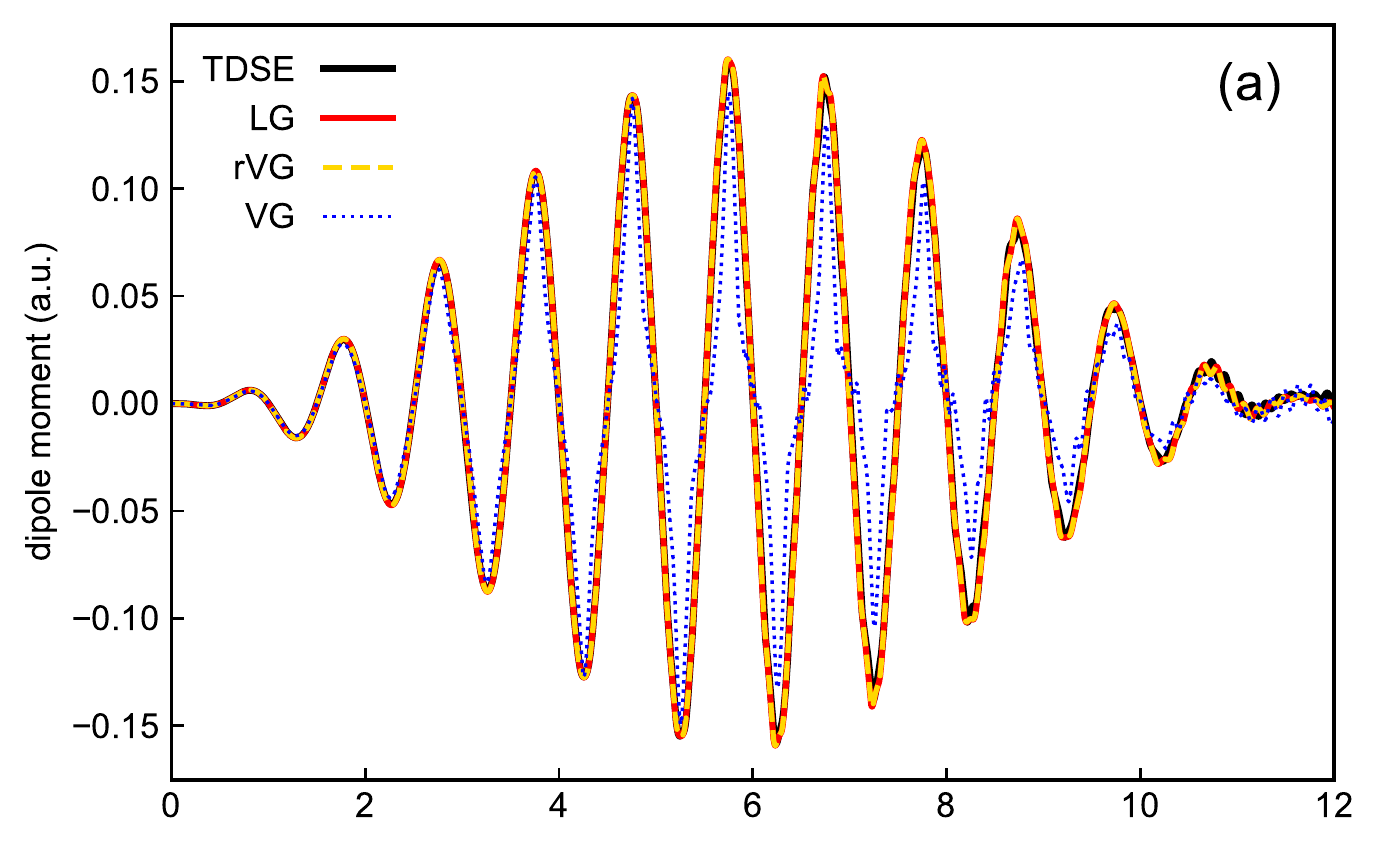}
    \end{minipage}
    %---------------------------------------------------%
    \begin{minipage}[htp]{1.0\hsize}
      \centering
      \includegraphics[width = \textwidth]{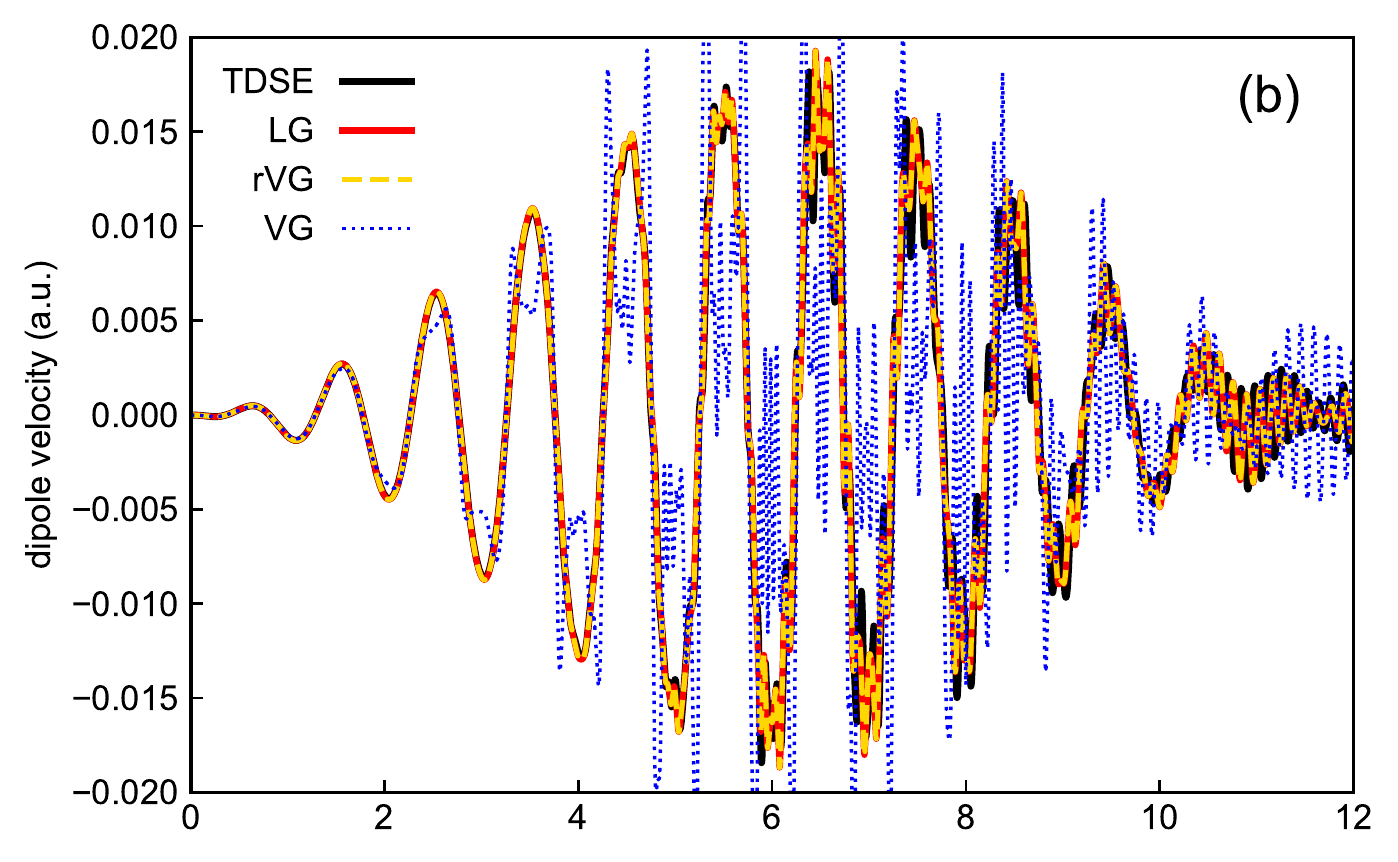}
    \end{minipage}
    %---------------------------------------------------%
    \begin{minipage}[htp]{1.0\hsize}
      \centering
      \includegraphics[width = \textwidth]{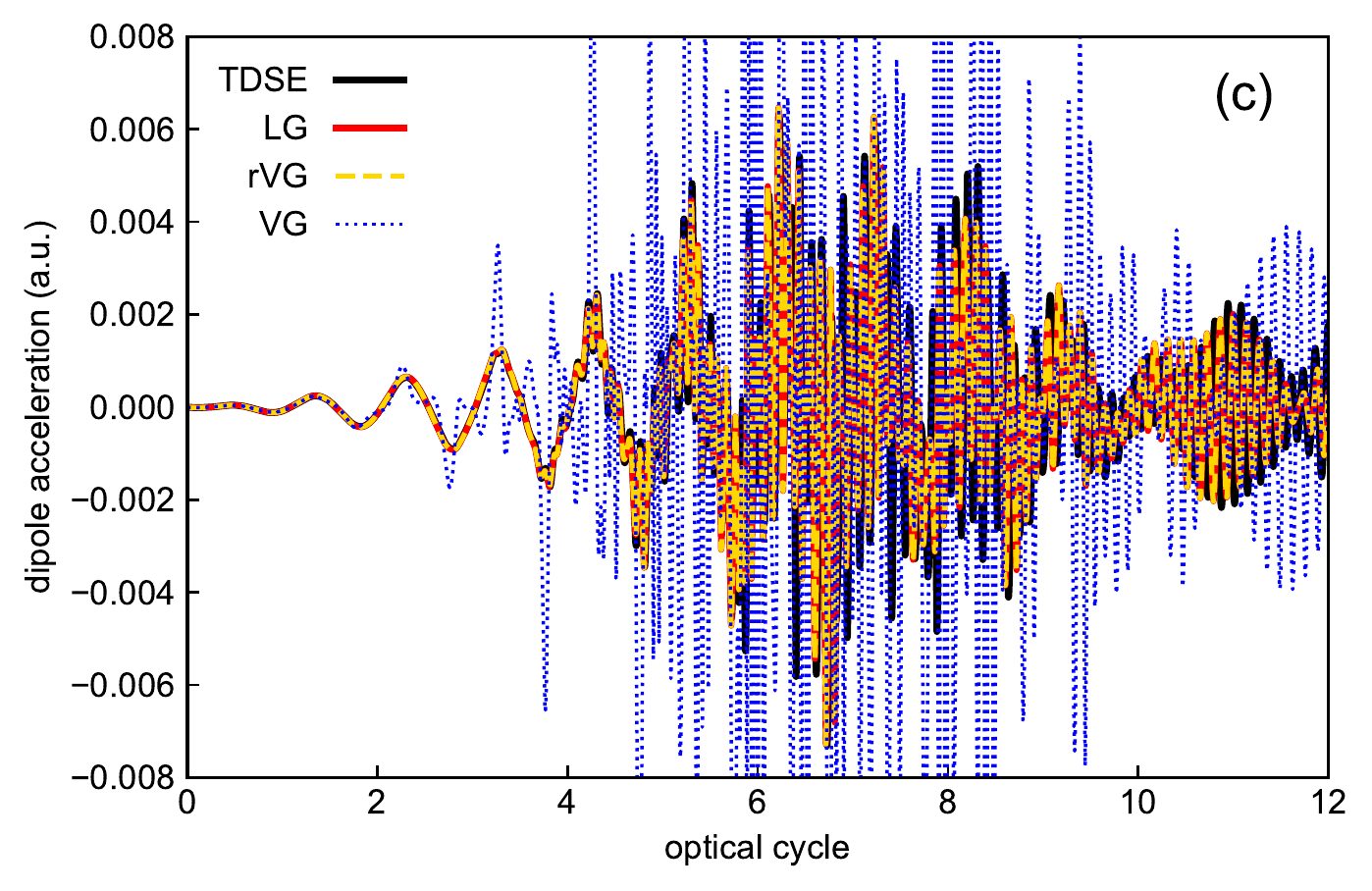}
      \caption{\label{He} Time evolution of (a) the dipole moment, (b) the dipole velocity, and (c) the dipole acceleration of He subject to a laser pulse with $\lambda = 400\ \text{nm}, I_0 = 4\times10^{14}\ \text{W/cm}^2$, and $N\down{opt} = 12$, obtained with the exact TDSE (courtesy of J. Burgd\"orfer) and the TDCIS method with length gauge (LG), conventional velocity gauge (VG), and rotated velocity gauge (rVG).}
    \end{minipage}
    %---------------------------------------------------%
\end{figure}

\begin{figure}
    %---------------------------------------------------%
    \begin{minipage}[htp]{1.0\hsize}
    \centering
    \includegraphics[clip, width = \textwidth]{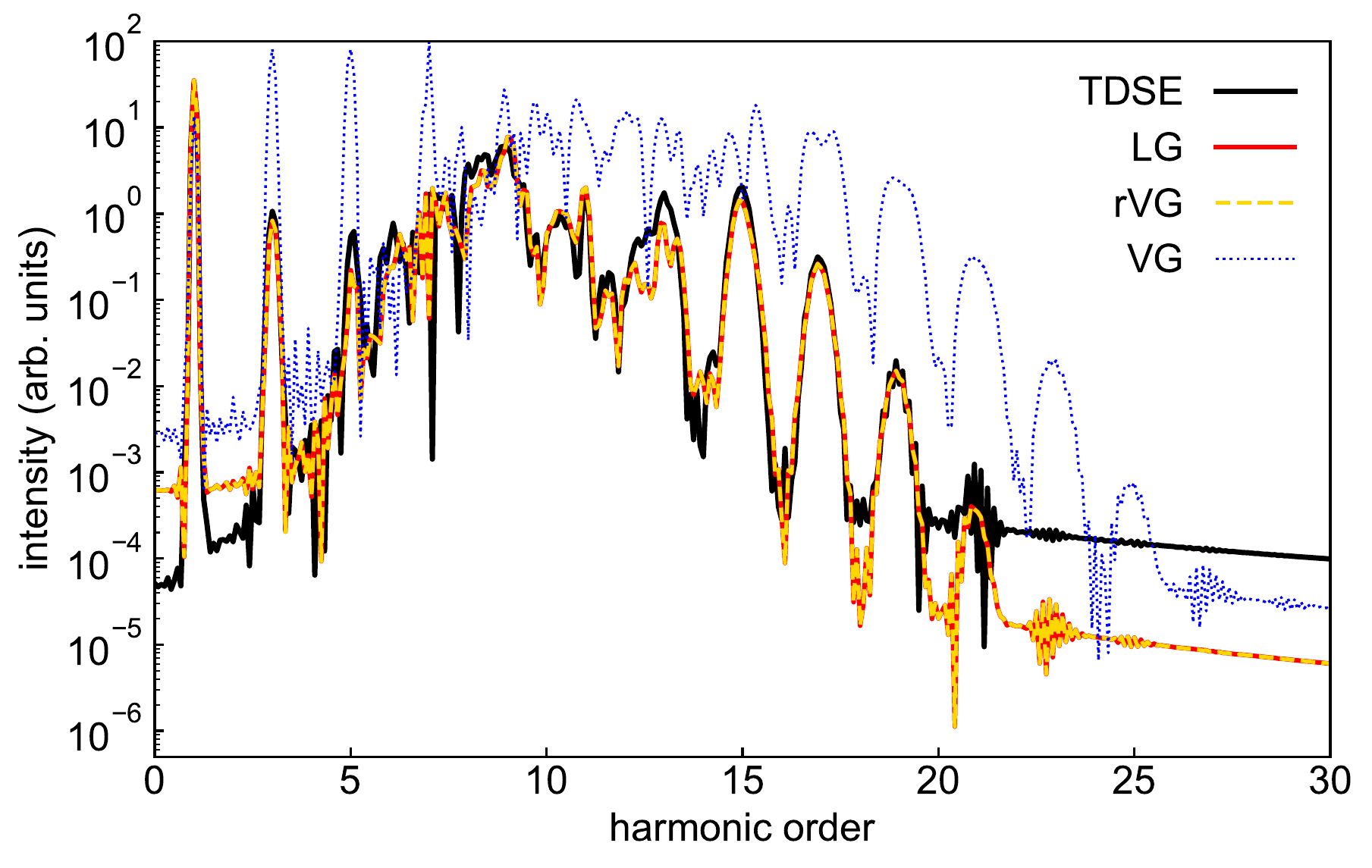}
      \caption{\label{He_hhg} HHG spectra from He exposed to a laser pulse with the same conditions as Fig. \ref{He}, computed from the dipole acceleration shown in Fig. \ref{He}. (c). Comparison of the TDSE result (courtesy of J. Burgd\"orfer) and TDCIS ones with LG, rVG, and VG.}
    \end{minipage}
    %---------------------------------------------------%
\end{figure}

Time evolution of the calculated dipole moment, dipole velocity, and dipole acceleration are shown in Fig.~\ref{He}, 
and HHG spectra obtained as the modulus squared of the Fourier transform of the dipole acceleration is presented in Fig~\ref{He_hhg}. In these figures, one can see the perfect agreement between the LG and rVG results, which numerically
confirms the gauge invariance between the two gauges.
In contrast, the results of conventional VG with fixed orbitals strongly deviate from them.
It should be noted that, from the comparison between LG (and rVG) and VG results alone, we cannot {\it a priori} tell which is more accurate.
The comparison with the TDSE results now reveals that the former reproduces the TDSE results much better than the latter, which convinces us of an {\it empirical} preference of the LG and rVG treatments.

We show the temporal evolution of the survival probability $P_0$ with $R\down{ion} = 20$ a.u. in Fig.~\ref{He_p0} (See Appendix for the $R\down{ion}$ dependence of $P_0$).
The conventional VG treatment strongly overestimates ionization.
Recalling that tunneling ionization is the first process of the three-step model \cite{PhysRevLett.71.1994,Piraux1993},
% high harmonic emission mainly occurs when an electron once escaped from the ion recombines with the parent ion.
% The strong overestimation of the ionization by the conventional VG treatment (underestimation of $P_0$ in Fig.~\ref{He_p0}) might lead to 
this explains the substantial overestimation of the HHG yield in Fig.~\ref{He_hhg}.
The LG and rVG results, on the other hand, underestimate tunneling ionization.
Correspondingly, indeed, we notice that the harmonic intensity is slightly underestimated in Fig.~\ref{He_hhg}.
% Through the comparison with numerically exact TDSE results for the full three dimensional He atom, we can confirm the
% conclusion that the rVG is more appropriate than the conventional VG for simulating the high-field phenomena, which 
% was previously demonstrated for a model 1D Hamiltonian \cite{Sato_Gauge_2018}. 

\begin{figure}
    %---------------------------------------------------%
    \begin{minipage}[htp]{1.0\hsize}
    \centering
      \includegraphics[clip, width = \textwidth]{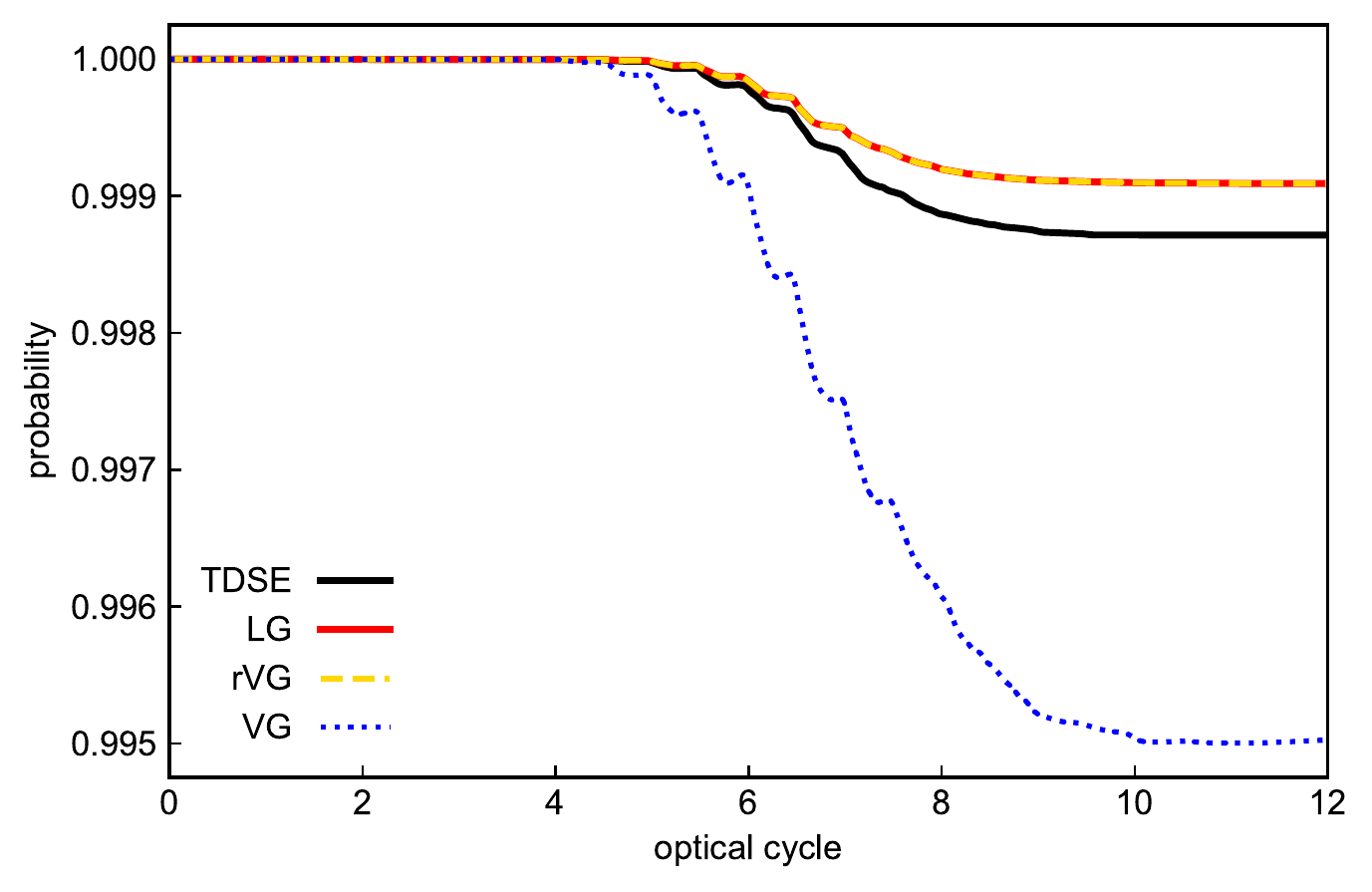}
      \caption{\label{He_p0} Time evolution of the survival probability $P_0$.    Comparison of the TDSE result (courtesy of J. Burgd\"orfer) and TDCIS ones with LG, rVG, and VG with $R\down{ion} = 20$\,a.u..
      } 
      \end{minipage}
  %---------------------------------------------------%
\end{figure}
%%%%%%%%%%%%%%%%%%%%%%%%%%%%%%%%%%%%%%%%%%%%%%%%%%%%%%%%%%%%
\subsection{\label{sec:Ne}Neon}
\begin{figure}[ht]
    %---------------------------------------------------%
    \begin{minipage}[htp]{1.0\hsize}
      \centering
      \includegraphics[width = \textwidth]{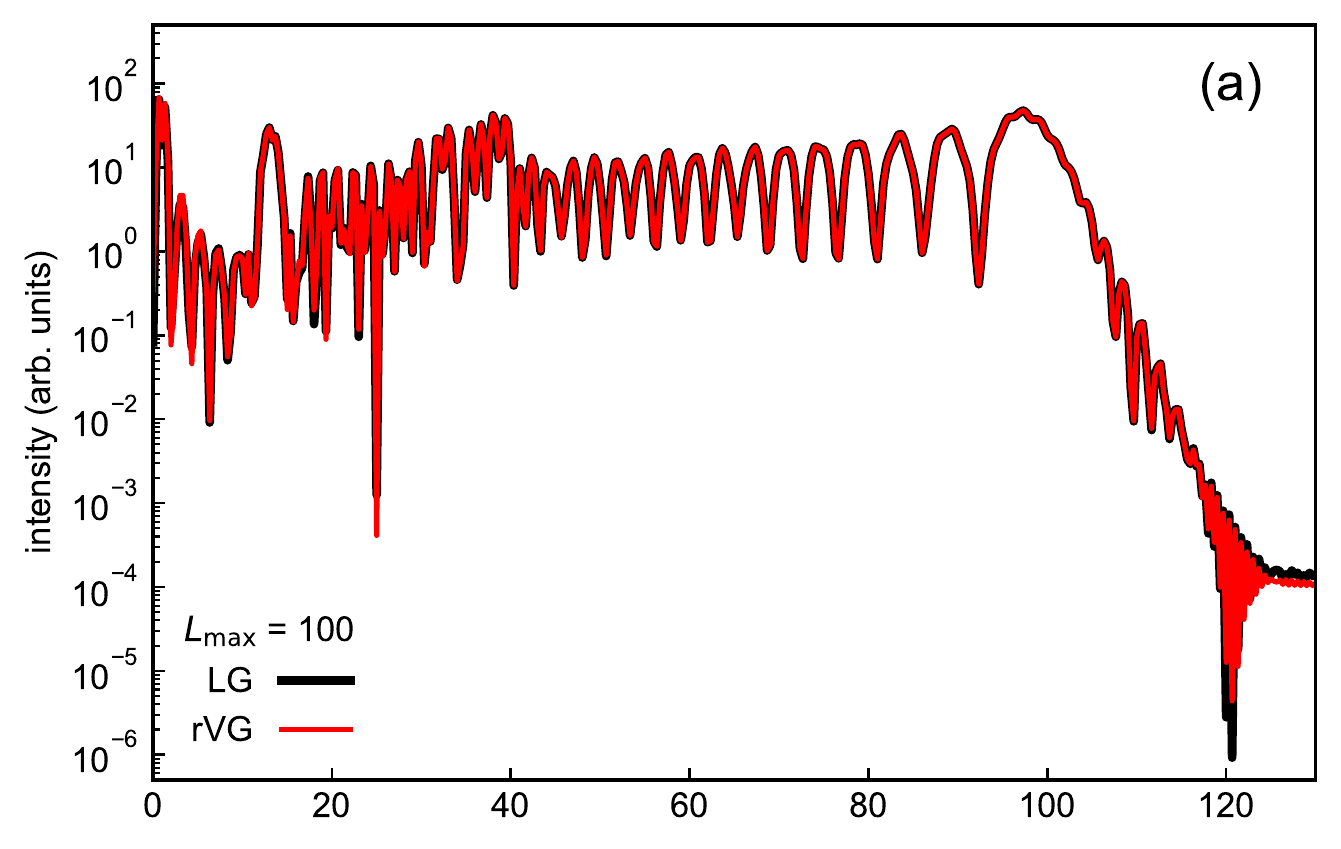}
    \end{minipage}
    %---------------------------------------------------%
    \begin{minipage}[htp]{1.0\hsize}
      \centering
      \includegraphics[width = \textwidth]{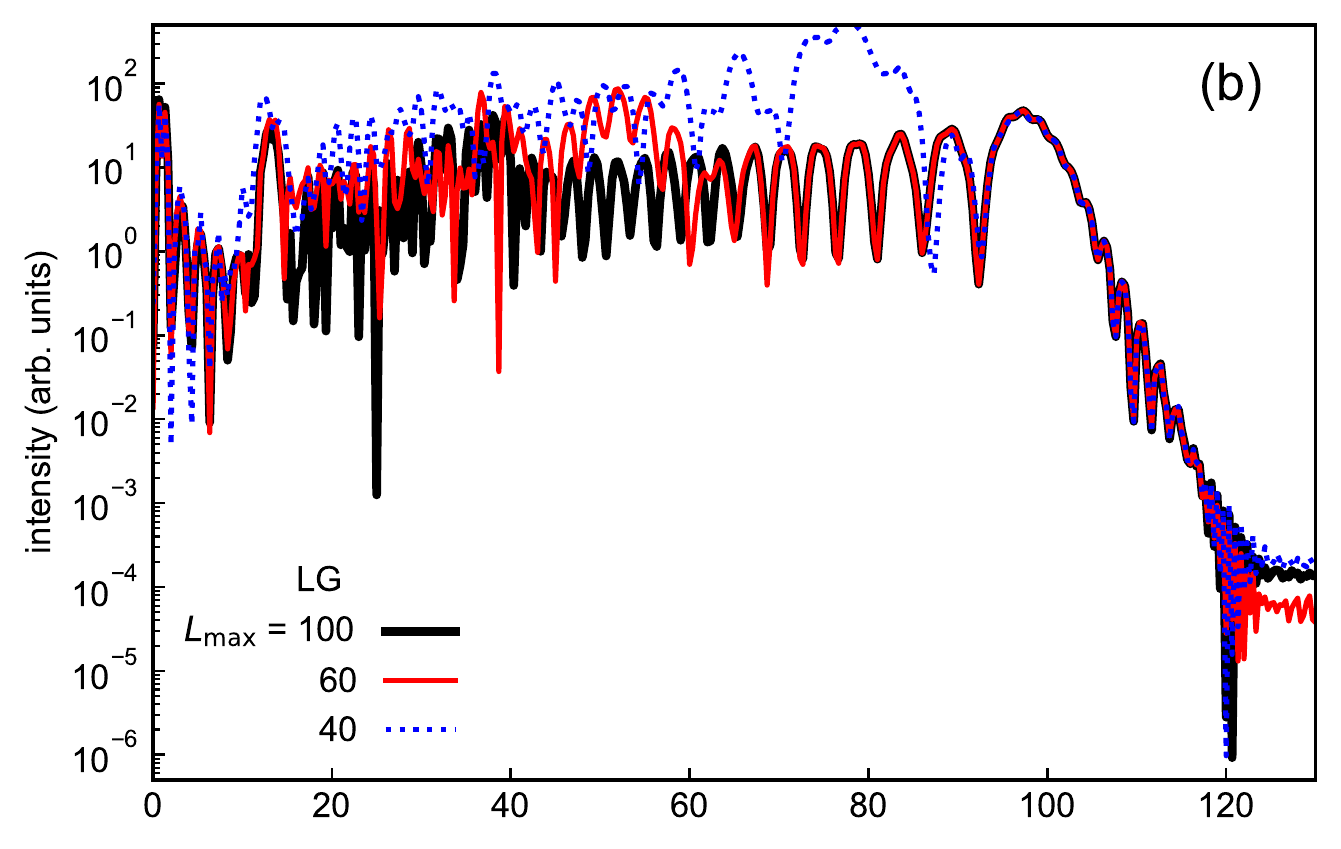}
    \end{minipage}
    %---------------------------------------------------%
    \begin{minipage}[htp]{1.0\hsize}
      \centering
      \includegraphics[width = \textwidth]{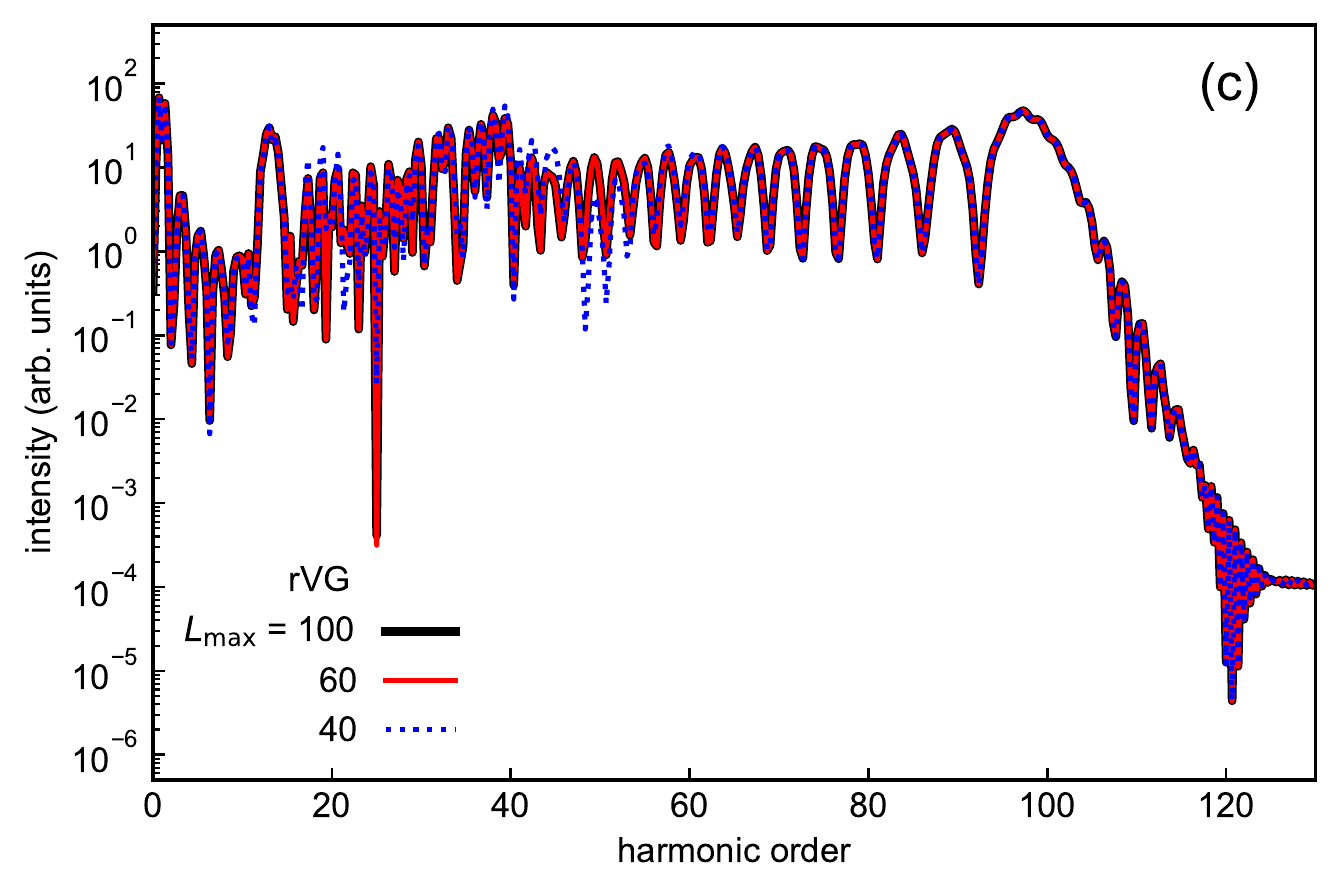}
      \caption{\label{Ne} HHG spectra of Ne subject to an IR laser pulse with a wavelength of 800 nm and an intensity of $1.0\times10^{14}\ \text{W/cm}^2$(a) Results of the LG and rVG with $L\down{max} = 100$. (b) Results of the LG with various $L\down{max}$. (c) Results of the rVG with various $L\down{max}$}
    \end{minipage}
    %---------------------------------------------------%
\end{figure}

We next consider a neon atom subject to a laser field with $\lambda =$ 800 nm, $I_0 = 8.0\times10^{14}$ W/cm$^2$, and $N\down{opt} = 3$ and discuss convergence with respect to the maximum angular momentum $L\down{max}$.
We show the HHG spectra calculated with various values of $L\down{max}$ in the LG and rVG in Fig.~\ref{Ne}.
Figure~\ref{Ne}.~(a) shows the equivalence between the LG and rVG for sufficiently large $L\down{max}(=100)$.
As can be seen in Fig.~\ref{Ne}.~(b), which shows LG results, $L\down{max} = 60$ is not sufficient to obtain a converged result.
%; $L\down{max} = 90$ or more is required for convergence.
On the other hand, the rVG requires far less $L\down{max}$; 
even $L\down{max} = 40$ well reproduces the result with $L\down{max} = 100$, and the spectrum is converged with $L\down{max} = 60$ [Fig.~\ref{Ne}.~(c)].
This observation indicates that the rVG TDCIS is simultaneously as accurate as the LG and as efficient as the VG.
%%%%%%%%%%%%%%%%%%%%%%%%%%%%%%%%%%%%%%%%%%%%%%%%%%%%%%%%%%%%
 \section{\label{sec:level5}conclusions}
We have presented a 3D numerical implementation of the recently formulated gauge-invariant TDCIS method \cite{Sato_Gauge_2018} for atoms subject to a linearly polarized intense laser field.
Compared to the conventional TDCIS method that uses CI coefficients as working variables, the present implementation introduces channel orbitals \cite{Rohringer_TDCIS_2006}, avoiding calculation and storage of numerous virtual orbitals.
% implemented TDCIS method has two remarkable features,
% absence of virtual orbitals and gauge invariance.
%Calculation and storage of numerous virtual orbitals is needed in the CI coefficient-based formulation of the TDCIS method.
%Converting the EOMs of the CI coefficients into those of channel orbitals gets rid of this demanding task and enables the TDCIS approach applicable to larger system like molecules.
%It is known that the velocity gauge enables efficient simulations of high-field phenomena compared to the length gauge.
%However, as a general consequence of the truncation of CI space, the conventional TDCIS method suffers from a violation of gauge invariance,
%which prevents efficient simulations of high-field phenomena within the velocity gauge.
%To enjoy the computational advantage of the velocity gauge with the TDCIS method,
%we have introduced the rVG which is connected with the LG by gauge transformation.
%We have successfully implemented this gauge-invariant TDCIS method to three-dimensional atoms with spherical harmonics expansion.
% Our implementation is the first application of the reformulated TDCIS method to three-dimensional systems.
We have applied this to He and Ne atoms, and calculated survival probabilities and HHG spectra for intense laser pulses.
% We applied the reformulated TDCIS method to helium and neon atom subject to an intense laser pulse.
The perfect agreement of the LG and rVG results obtained with a sufficiently large number of partial waves numerically demonstrates the gauge invariance of the method.
The comparison with the numerically exact TDSE results for He shows the rVG and LG's superiority to the conventional VG in terms of accuracy.
The VG largely overestimates tunneling ionization and, then, harmonic intensity.
% spectrum, as a result of the overestimation of the ionization.
The analysis with neon reveals that the rVG has advantage in computational efficiency over the LG in terms of the number of spherical harmonics required to obtain converged HHG spectrum.
Thus, our gauge-invariant reformulation will make TDCIS a promising approach for multielectron dynamics not only in atoms but also in molecules driven by high-intensity laser fields.

\section*{\label{sec:level6}acknowledgments}
We are grateful to Prof.~J.~Burgd\"orfer of Vienna University of Technology for providing us with He TDSE benchmark results.
This research was supported in part by a Grant-in-Aid for
Scientific Research (Grants No. 16H03881, No. 17K05070,
No. 18H03891, and No. 19H00869) from the Ministry of Education, Culture,
Sports, Science and Technology (MEXT) of Japan. 
This research was also partially supported by JST COI (Grant No.~JPMJCE1313), JST CREST (Grant No.~JPMJCR15N1), and by Quantum Leap Flagship Program of MEXT.
T. T. gratefully acknowledges support from the Graduate School of Engineering, The University of Tokyo, Doctoral Student Special Incentives Program (SEUT Fellowship).

\appendix*
\section{$R\down{ion}$ dependence on $P_0$}
\begin{figure}[ht]
    \begin{minipage}[htp]{1.0\hsize}
    \centering
    \includegraphics[clip, width = \textwidth]{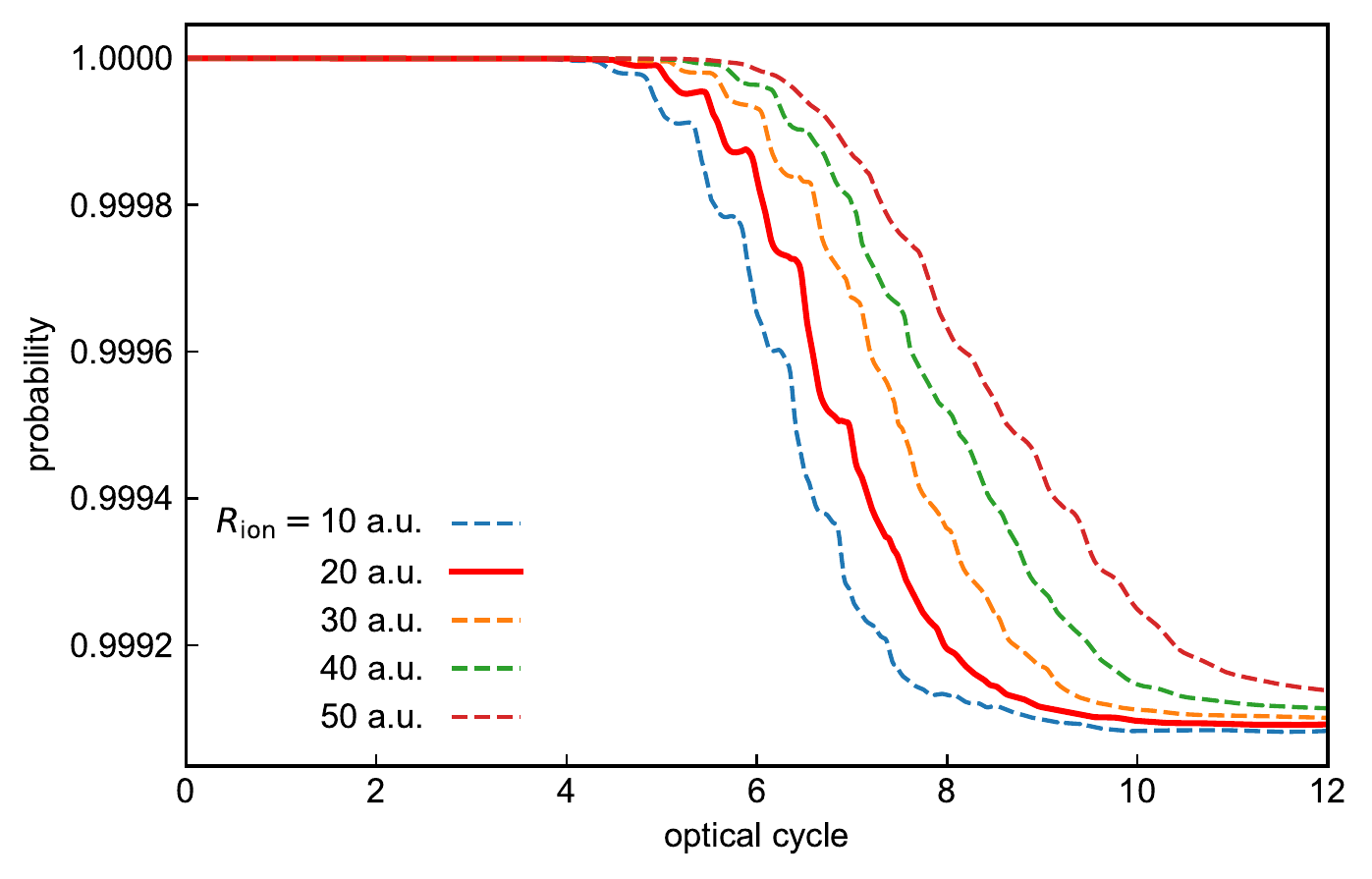}
    \caption{\label{He_p0_rion}Temporal evolution of $P_0$ for various values of $R\down{ion}$, extracted from LG TDCIS results.}
    \end{minipage}
 \end{figure}

 We show $P_0$ from the LG TDCIS results with various values of $R\down{ion}$ in fig.~\ref{He_p0_rion}.
We can see stepwise ejection of electron wavepackets that propagate outwards; the larger $R\down{ion}$ is, the later $P_0$ is depleted.
%  This figure represents the procedure electric wave packet goes away from the parent ion.
%  The larger $R\down{ion}$ is, the longer time ejected electron spends inside the inner region, and it results in the slow depletion of $P_0$ in large $R\down{ion}$.
 In the cases of $R\down{ion} =$ 10, 20, and 30 a.u.,
$P_0$ nearly reaches approximately the same final value in the end of the simlation (after twelve optical cycles).
%  one can see stepwise decrees of $P_0$,
%  which is a characteristic of tunnel ionization.
%  In these cases, $P_0$ remains constant after $t = 10$ optical cycles,
%  which can be interpreted as a spatial separation of the ejected electron and its parent ion.
%  Although $P_0$ at the end of the pulse depends on $R\down{ion}$,
%  this dependence is insignificant in the case of $R\down{ion} = 10\sim30$ a.u..

%In the ETDRK4, $\vb*{h}$ does not have time-dependence.
%\bibliography{ref}% Produces the bibliography via BibTeX.
\input{main.bbl}
\end{document}

%% file: main.bbl

%% file: main.bbl
\begin{thebibliography}{53}%
\makeatletter
\providecommand \@ifxundefined [1]{%
 \@ifx{#1\undefined}
}%
\providecommand \@ifnum [1]{%
 \ifnum #1\expandafter \@firstoftwo
 \else \expandafter \@secondoftwo
 \fi
}%
\providecommand \@ifx [1]{%
 \ifx #1\expandafter \@firstoftwo
 \else \expandafter \@secondoftwo
 \fi
}%
\providecommand \natexlab [1]{#1}%
\providecommand \enquote  [1]{``#1''}%
\providecommand \bibnamefont  [1]{#1}%
\providecommand \bibfnamefont [1]{#1}%
\providecommand \citenamefont [1]{#1}%
\providecommand \href@noop [0]{\@secondoftwo}%
\providecommand \href [0]{\begingroup \@sanitize@url \@href}%
\providecommand \@href[1]{\@@startlink{#1}\@@href}%
\providecommand \@@href[1]{\endgroup#1\@@endlink}%
\providecommand \@sanitize@url [0]{\catcode `\\12\catcode `\$12\catcode
  `\&12\catcode `\#12\catcode `\^12\catcode `\_12\catcode `\%12\relax}%
\providecommand \@@startlink[1]{}%
\providecommand \@@endlink[0]{}%
\providecommand \url  [0]{\begingroup\@sanitize@url \@url }%
\providecommand \@url [1]{\endgroup\@href {#1}{\urlprefix }}%
\providecommand \urlprefix  [0]{URL }%
\providecommand \Eprint [0]{\href }%
\providecommand \doibase [0]{http://dx.doi.org/}%
\providecommand \selectlanguage [0]{\@gobble}%
\providecommand \bibinfo  [0]{\@secondoftwo}%
\providecommand \bibfield  [0]{\@secondoftwo}%
\providecommand \translation [1]{[#1]}%
\providecommand \BibitemOpen [0]{}%
\providecommand \bibitemStop [0]{}%
\providecommand \bibitemNoStop [0]{.\EOS\space}%
\providecommand \EOS [0]{\spacefactor3000\relax}%
\providecommand \BibitemShut  [1]{\csname bibitem#1\endcsname}%
\let\auto@bib@innerbib\@empty
%</preamble>
\bibitem [{\citenamefont {Krausz}\ and\ \citenamefont
  {Ivanov}(2009)}]{RevModPhys.81.163}%
  \BibitemOpen
  \bibfield  {author} {\bibinfo {author} {\bibfnamefont {F.}~\bibnamefont
  {Krausz}}\ and\ \bibinfo {author} {\bibfnamefont {M.}~\bibnamefont
  {Ivanov}},\ }\href {\doibase 10.1103/RevModPhys.81.163} {\bibfield  {journal}
  {\bibinfo  {journal} {Rev. Mod. Phys.}\ }\textbf {\bibinfo {volume} {81}},\
  \bibinfo {pages} {163} (\bibinfo {year} {2009})}\BibitemShut {NoStop}%
\bibitem [{\citenamefont {Nisoli}\ \emph {et~al.}(2017)\citenamefont {Nisoli},
  \citenamefont {Decleva}, \citenamefont {Calegari}, \citenamefont {Palacios},\
  and\ \citenamefont {Mart{\'\i}n}}]{Chem.Rev.6b00453}%
  \BibitemOpen
  \bibfield  {author} {\bibinfo {author} {\bibfnamefont {M.}~\bibnamefont
  {Nisoli}}, \bibinfo {author} {\bibfnamefont {P.}~\bibnamefont {Decleva}},
  \bibinfo {author} {\bibfnamefont {F.}~\bibnamefont {Calegari}}, \bibinfo
  {author} {\bibfnamefont {A.}~\bibnamefont {Palacios}}, \ and\ \bibinfo
  {author} {\bibfnamefont {F.}~\bibnamefont {Mart{\'\i}n}},\ }\href {\doibase
  10.1021/acs.chemrev.6b00453} {\bibfield  {journal} {\bibinfo  {journal}
  {Chem. Rev.}\ }\textbf {\bibinfo {volume} {117}},\ \bibinfo {pages} {10760}
  (\bibinfo {year} {2017})}\BibitemShut {NoStop}%
\bibitem [{\citenamefont {Gallmann}\ \emph {et~al.}(2013)\citenamefont
  {Gallmann}, \citenamefont {Cirelli},\ and\ \citenamefont
  {Keller}}]{Annu.Rev.Phys.Chem.2013}%
  \BibitemOpen
  \bibfield  {author} {\bibinfo {author} {\bibfnamefont {L.}~\bibnamefont
  {Gallmann}}, \bibinfo {author} {\bibfnamefont {C.}~\bibnamefont {Cirelli}}, \
  and\ \bibinfo {author} {\bibfnamefont {U.}~\bibnamefont {Keller}},\
  }\href@noop {} {\bibfield  {journal} {\bibinfo  {journal} {Annu. Rev. Phys.
  Chem.}\ }\textbf {\bibinfo {volume} {63}},\ \bibinfo {pages} {447} (\bibinfo
  {year} {2013})}\BibitemShut {NoStop}%
\bibitem [{\citenamefont {Zanghellini}\ \emph {et~al.}(2003)\citenamefont
  {Zanghellini}, \citenamefont {Kitzler}, \citenamefont {Fabian}, \citenamefont
  {Brabec},\ and\ \citenamefont {Scrinzi}}]{Zanghellini2003}%
  \BibitemOpen
  \bibfield  {author} {\bibinfo {author} {\bibfnamefont {J.}~\bibnamefont
  {Zanghellini}}, \bibinfo {author} {\bibfnamefont {M.}~\bibnamefont
  {Kitzler}}, \bibinfo {author} {\bibfnamefont {C.}~\bibnamefont {Fabian}},
  \bibinfo {author} {\bibfnamefont {T.}~\bibnamefont {Brabec}}, \ and\ \bibinfo
  {author} {\bibfnamefont {A.}~\bibnamefont {Scrinzi}},\ }\href
  {http://www.maik.ru/full/lasphys/03/8/lasphys8{\_}03p1064full.pdf} {\bibfield
   {journal} {\bibinfo  {journal} {Laser Phys.}\ }\textbf {\bibinfo {volume}
  {13}},\ \bibinfo {pages} {1064} (\bibinfo {year} {2003})}\BibitemShut
  {NoStop}%
\bibitem [{\citenamefont {Kato}\ and\ \citenamefont {Kono}(2004)}]{Kato2004}%
  \BibitemOpen
  \bibfield  {author} {\bibinfo {author} {\bibfnamefont {T.}~\bibnamefont
  {Kato}}\ and\ \bibinfo {author} {\bibfnamefont {H.}~\bibnamefont {Kono}},\
  }\href {\doibase 10.1016/J.CPLETT.2004.05.106} {\bibfield  {journal}
  {\bibinfo  {journal} {Chem. Phys. Lett.}\ }\textbf {\bibinfo {volume}
  {392}},\ \bibinfo {pages} {533} (\bibinfo {year} {2004})}\BibitemShut
  {NoStop}%
\bibitem [{\citenamefont {Haxton}\ and\ \citenamefont
  {McCurdy}(2015{\natexlab{a}})}]{PhysRevA.91.012509}%
  \BibitemOpen
  \bibfield  {author} {\bibinfo {author} {\bibfnamefont {D.~J.}\ \bibnamefont
  {Haxton}}\ and\ \bibinfo {author} {\bibfnamefont {C.~W.}\ \bibnamefont
  {McCurdy}},\ }\href {\doibase 10.1103/PhysRevA.91.012509} {\bibfield
  {journal} {\bibinfo  {journal} {Phys. Rev. A}\ }\textbf {\bibinfo {volume}
  {91}},\ \bibinfo {pages} {012509} (\bibinfo {year}
  {2015}{\natexlab{a}})}\BibitemShut {NoStop}%
\bibitem [{\citenamefont {Miyagi}\ and\ \citenamefont
  {Madsen}(2014{\natexlab{a}})}]{PhysRevA.89.063416}%
  \BibitemOpen
  \bibfield  {author} {\bibinfo {author} {\bibfnamefont {H.}~\bibnamefont
  {Miyagi}}\ and\ \bibinfo {author} {\bibfnamefont {L.~B.}\ \bibnamefont
  {Madsen}},\ }\href {\doibase 10.1103/PhysRevA.89.063416} {\bibfield
  {journal} {\bibinfo  {journal} {Phys. Rev. A}\ }\textbf {\bibinfo {volume}
  {89}},\ \bibinfo {pages} {063416} (\bibinfo {year}
  {2014}{\natexlab{a}})}\BibitemShut {NoStop}%
\bibitem [{\citenamefont {Caillat}\ \emph {et~al.}(2005)\citenamefont
  {Caillat}, \citenamefont {Zanghellini}, \citenamefont {Kitzler},
  \citenamefont {Koch}, \citenamefont {Kreuzer},\ and\ \citenamefont
  {Scrinzi}}]{Caillat_MCTDHF}%
  \BibitemOpen
  \bibfield  {author} {\bibinfo {author} {\bibfnamefont {J.}~\bibnamefont
  {Caillat}}, \bibinfo {author} {\bibfnamefont {J.}~\bibnamefont
  {Zanghellini}}, \bibinfo {author} {\bibfnamefont {M.}~\bibnamefont
  {Kitzler}}, \bibinfo {author} {\bibfnamefont {O.}~\bibnamefont {Koch}},
  \bibinfo {author} {\bibfnamefont {W.}~\bibnamefont {Kreuzer}}, \ and\
  \bibinfo {author} {\bibfnamefont {A.}~\bibnamefont {Scrinzi}},\ }\href
  {\doibase 10.1103/PhysRevA.71.012712} {\bibfield  {journal} {\bibinfo
  {journal} {Phys. Rev. A}\ }\textbf {\bibinfo {volume} {71}},\ \bibinfo
  {pages} {012712} (\bibinfo {year} {2005})}\BibitemShut {NoStop}%
\bibitem [{\citenamefont {Miyagi}\ and\ \citenamefont
  {Madsen}(2013{\natexlab{a}})}]{PhysRevA.87.062511}%
  \BibitemOpen
  \bibfield  {author} {\bibinfo {author} {\bibfnamefont {H.}~\bibnamefont
  {Miyagi}}\ and\ \bibinfo {author} {\bibfnamefont {L.~B.}\ \bibnamefont
  {Madsen}},\ }\href {\doibase 10.1103/PhysRevA.87.062511} {\bibfield
  {journal} {\bibinfo  {journal} {Phys. Rev. A}\ }\textbf {\bibinfo {volume}
  {87}},\ \bibinfo {pages} {062511} (\bibinfo {year}
  {2013}{\natexlab{a}})}\BibitemShut {NoStop}%
\bibitem [{\citenamefont {Miyagi}\ and\ \citenamefont
  {Madsen}(2013{\natexlab{b}})}]{Miyagi_2013}%
  \BibitemOpen
  \bibfield  {author} {\bibinfo {author} {\bibfnamefont {H.}~\bibnamefont
  {Miyagi}}\ and\ \bibinfo {author} {\bibfnamefont {L.}~\bibnamefont
  {Madsen}},\ }\href@noop {} {\bibfield  {journal} {\bibinfo  {journal} {Phys.
  Rev. A}\ }\textbf {\bibinfo {volume} {87}},\ \bibinfo {pages} {062511}
  (\bibinfo {year} {2013}{\natexlab{b}})}\BibitemShut {NoStop}%
\bibitem [{\citenamefont {Sato}\ and\ \citenamefont
  {Ishikawa}(2013)}]{Sato_CAS_2013}%
  \BibitemOpen
  \bibfield  {author} {\bibinfo {author} {\bibfnamefont {T.}~\bibnamefont
  {Sato}}\ and\ \bibinfo {author} {\bibfnamefont {K.~L.}\ \bibnamefont
  {Ishikawa}},\ }\href {\doibase 10.1103/PhysRevA.88.023402} {\bibfield
  {journal} {\bibinfo  {journal} {Phys. Rev. A}\ }\textbf {\bibinfo {volume}
  {88}},\ \bibinfo {pages} {023402} (\bibinfo {year} {2013})}\BibitemShut
  {NoStop}%
\bibitem [{\citenamefont {Miyagi}\ and\ \citenamefont
  {Madsen}(2014{\natexlab{b}})}]{Miyagi_2014}%
  \BibitemOpen
  \bibfield  {author} {\bibinfo {author} {\bibfnamefont {H.}~\bibnamefont
  {Miyagi}}\ and\ \bibinfo {author} {\bibfnamefont {L.~B.}\ \bibnamefont
  {Madsen}},\ }\href@noop {} {\bibfield  {journal} {\bibinfo  {journal} {Phys.
  Rev. A}\ }\textbf {\bibinfo {volume} {89}},\ \bibinfo {pages} {063416}
  (\bibinfo {year} {2014}{\natexlab{b}})}\BibitemShut {NoStop}%
\bibitem [{\citenamefont {Haxton}\ and\ \citenamefont
  {McCurdy}(2015{\natexlab{b}})}]{Haxton_2015}%
  \BibitemOpen
  \bibfield  {author} {\bibinfo {author} {\bibfnamefont {D.~J.}\ \bibnamefont
  {Haxton}}\ and\ \bibinfo {author} {\bibfnamefont {C.~W.}\ \bibnamefont
  {McCurdy}},\ }\href@noop {} {\bibfield  {journal} {\bibinfo  {journal} {Phys.
  Rev. A}\ }\textbf {\bibinfo {volume} {91}},\ \bibinfo {pages} {012509}
  (\bibinfo {year} {2015}{\natexlab{b}})}\BibitemShut {NoStop}%
\bibitem [{\citenamefont {Sato}\ and\ \citenamefont
  {Ishikawa}(2015)}]{Sato_ORMAS_2015}%
  \BibitemOpen
  \bibfield  {author} {\bibinfo {author} {\bibfnamefont {T.}~\bibnamefont
  {Sato}}\ and\ \bibinfo {author} {\bibfnamefont {K.~L.}\ \bibnamefont
  {Ishikawa}},\ }\href {\doibase 10.1103/PhysRevA.91.023417} {\bibfield
  {journal} {\bibinfo  {journal} {Phys. Rev. A}\ }\textbf {\bibinfo {volume}
  {91}},\ \bibinfo {pages} {023417} (\bibinfo {year} {2015})}\BibitemShut
  {NoStop}%
\bibitem [{\citenamefont {Sato}\ \emph {et~al.}(2016)\citenamefont {Sato},
  \citenamefont {Ishikawa}, \citenamefont {B\ifmmode~\check{r}\else
  \v{r}\fi{}ezinov\'a}, \citenamefont {Lackner}, \citenamefont {Nagele},\ and\
  \citenamefont {Burgd\"orfer}}]{Sato_CAS2_2016}%
  \BibitemOpen
  \bibfield  {author} {\bibinfo {author} {\bibfnamefont {T.}~\bibnamefont
  {Sato}}, \bibinfo {author} {\bibfnamefont {K.~L.}\ \bibnamefont {Ishikawa}},
  \bibinfo {author} {\bibfnamefont {I.}~\bibnamefont {B\ifmmode~\check{r}\else
  \v{r}\fi{}ezinov\'a}}, \bibinfo {author} {\bibfnamefont {F.}~\bibnamefont
  {Lackner}}, \bibinfo {author} {\bibfnamefont {S.}~\bibnamefont {Nagele}}, \
  and\ \bibinfo {author} {\bibfnamefont {J.}~\bibnamefont {Burgd\"orfer}},\
  }\href {\doibase 10.1103/PhysRevA.94.023405} {\bibfield  {journal} {\bibinfo
  {journal} {Phys. Rev. A}\ }\textbf {\bibinfo {volume} {94}},\ \bibinfo
  {pages} {023405} (\bibinfo {year} {2016})}\BibitemShut {NoStop}%
\bibitem [{\citenamefont {Anzaki}\ \emph {et~al.}(2017)\citenamefont {Anzaki},
  \citenamefont {Sato},\ and\ \citenamefont {Ishikawa}}]{Anzaki_MCSCF_2017}%
  \BibitemOpen
  \bibfield  {author} {\bibinfo {author} {\bibfnamefont {R.}~\bibnamefont
  {Anzaki}}, \bibinfo {author} {\bibfnamefont {T.}~\bibnamefont {Sato}}, \ and\
  \bibinfo {author} {\bibfnamefont {K.~L.}\ \bibnamefont {Ishikawa}},\ }\href
  {\doibase 10.1039/C7CP02086D} {\bibfield  {journal} {\bibinfo  {journal}
  {Phys. Chem. Chem. Phys.}\ }\textbf {\bibinfo {volume} {19}},\ \bibinfo
  {pages} {22008} (\bibinfo {year} {2017})}\BibitemShut {NoStop}%
\bibitem [{\citenamefont {Kvaal}(2012)}]{Kvaal_2012}%
  \BibitemOpen
  \bibfield  {author} {\bibinfo {author} {\bibfnamefont {S.}~\bibnamefont
  {Kvaal}},\ }\href@noop {} {\bibfield  {journal} {\bibinfo  {journal} {J.
  Chem. Phys.}\ }\textbf {\bibinfo {volume} {136}},\ \bibinfo {pages} {194109}
  (\bibinfo {year} {2012})}\BibitemShut {NoStop}%
\bibitem [{\citenamefont {Sato}\ \emph
  {et~al.}(2018{\natexlab{a}})\citenamefont {Sato}, \citenamefont {Pathak},
  \citenamefont {Orimo},\ and\ \citenamefont {Ishikawa}}]{Sato_OCC_2018}%
  \BibitemOpen
  \bibfield  {author} {\bibinfo {author} {\bibfnamefont {T.}~\bibnamefont
  {Sato}}, \bibinfo {author} {\bibfnamefont {H.}~\bibnamefont {Pathak}},
  \bibinfo {author} {\bibfnamefont {Y.}~\bibnamefont {Orimo}}, \ and\ \bibinfo
  {author} {\bibfnamefont {K.~L.}\ \bibnamefont {Ishikawa}},\ }\href {\doibase
  10.1063/1.5020633} {\bibfield  {journal} {\bibinfo  {journal} {J. Chem.
  Phys.}\ }\textbf {\bibinfo {volume} {148}},\ \bibinfo {pages} {051101}
  (\bibinfo {year} {2018}{\natexlab{a}})}\BibitemShut {NoStop}%
\bibitem [{\citenamefont {Lysaght}\ \emph {et~al.}(2009)\citenamefont
  {Lysaght}, \citenamefont {van~der Hart},\ and\ \citenamefont
  {Burke}}]{PhysRevA.79.053411}%
  \BibitemOpen
  \bibfield  {author} {\bibinfo {author} {\bibfnamefont {M.~A.}\ \bibnamefont
  {Lysaght}}, \bibinfo {author} {\bibfnamefont {H.~W.}\ \bibnamefont {van~der
  Hart}}, \ and\ \bibinfo {author} {\bibfnamefont {P.~G.}\ \bibnamefont
  {Burke}},\ }\href {\doibase 10.1103/PhysRevA.79.053411} {\bibfield  {journal}
  {\bibinfo  {journal} {Phys. Rev. A}\ }\textbf {\bibinfo {volume} {79}},\
  \bibinfo {pages} {053411} (\bibinfo {year} {2009})}\BibitemShut {NoStop}%
\bibitem [{\citenamefont {Lysaght}\ \emph {et~al.}(2008)\citenamefont
  {Lysaght}, \citenamefont {Burke},\ and\ \citenamefont {van~der
  Hart}}]{PhysRevLett.101.253001}%
  \BibitemOpen
  \bibfield  {author} {\bibinfo {author} {\bibfnamefont {M.~A.}\ \bibnamefont
  {Lysaght}}, \bibinfo {author} {\bibfnamefont {P.~G.}\ \bibnamefont {Burke}},
  \ and\ \bibinfo {author} {\bibfnamefont {H.~W.}\ \bibnamefont {van~der
  Hart}},\ }\href {\doibase 10.1103/PhysRevLett.101.253001} {\bibfield
  {journal} {\bibinfo  {journal} {Phys. Rev. Lett.}\ }\textbf {\bibinfo
  {volume} {101}},\ \bibinfo {pages} {253001} (\bibinfo {year}
  {2008})}\BibitemShut {NoStop}%
\bibitem [{\citenamefont {Burke}\ and\ \citenamefont
  {Burke}(1997)}]{Burke_1997}%
  \BibitemOpen
  \bibfield  {author} {\bibinfo {author} {\bibfnamefont {P.~G.}\ \bibnamefont
  {Burke}}\ and\ \bibinfo {author} {\bibfnamefont {V.~M.}\ \bibnamefont
  {Burke}},\ }\href {\doibase 10.1088/0953-4075/30/11/002} {\bibfield
  {journal} {\bibinfo  {journal} {J. Phys. B: At., Mol. Opt. Phys.}\ }\textbf
  {\bibinfo {volume} {30}},\ \bibinfo {pages} {L383} (\bibinfo {year}
  {1997})}\BibitemShut {NoStop}%
\bibitem [{\citenamefont {Moore}\ \emph {et~al.}(2011)\citenamefont {Moore},
  \citenamefont {Lysaght}, \citenamefont {Nikolopoulos}, \citenamefont
  {Parker}, \citenamefont {{Van Der Hart}},\ and\ \citenamefont
  {Taylor}}]{Moore2011}%
  \BibitemOpen
  \bibfield  {author} {\bibinfo {author} {\bibfnamefont {L.~R.}\ \bibnamefont
  {Moore}}, \bibinfo {author} {\bibfnamefont {M.~A.}\ \bibnamefont {Lysaght}},
  \bibinfo {author} {\bibfnamefont {L.~A.}\ \bibnamefont {Nikolopoulos}},
  \bibinfo {author} {\bibfnamefont {J.~S.}\ \bibnamefont {Parker}}, \bibinfo
  {author} {\bibfnamefont {H.~W.}\ \bibnamefont {{Van Der Hart}}}, \ and\
  \bibinfo {author} {\bibfnamefont {K.~T.}\ \bibnamefont {Taylor}},\ }\href
  {\doibase 10.1080/09500340.2011.559315} {\bibfield  {journal} {\bibinfo
  {journal} {J. Mod. Opt.}\ }\textbf {\bibinfo {volume} {58}},\ \bibinfo
  {pages} {1132} (\bibinfo {year} {2011})}\BibitemShut {NoStop}%
\bibitem [{\citenamefont {Clarke}\ \emph {et~al.}(2018)\citenamefont {Clarke},
  \citenamefont {Armstrong}, \citenamefont {Brown},\ and\ \citenamefont {{Van
  Der Hart}}}]{Clarke2018}%
  \BibitemOpen
  \bibfield  {author} {\bibinfo {author} {\bibfnamefont {D.~D.}\ \bibnamefont
  {Clarke}}, \bibinfo {author} {\bibfnamefont {G.~S.}\ \bibnamefont
  {Armstrong}}, \bibinfo {author} {\bibfnamefont {A.~C.}\ \bibnamefont
  {Brown}}, \ and\ \bibinfo {author} {\bibfnamefont {H.~W.}\ \bibnamefont {{Van
  Der Hart}}},\ }\href {\doibase 10.1103/PhysRevA.98.053442} {\bibfield
  {journal} {\bibinfo  {journal} {Phys. Rev. A}\ }\textbf {\bibinfo {volume}
  {98}},\ \bibinfo {pages} {1} (\bibinfo {year} {2018})}\BibitemShut {NoStop}%
\bibitem [{\citenamefont {Lackner}\ \emph {et~al.}(2015)\citenamefont
  {Lackner}, \citenamefont {B\ifmmode~\check{r}\else \v{r}\fi{}ezinov\'a},
  \citenamefont {Sato}, \citenamefont {Ishikawa},\ and\ \citenamefont
  {Burgd\"orfer}}]{PhysRevA.91.023412}%
  \BibitemOpen
  \bibfield  {author} {\bibinfo {author} {\bibfnamefont {F.}~\bibnamefont
  {Lackner}}, \bibinfo {author} {\bibfnamefont {I.}~\bibnamefont
  {B\ifmmode~\check{r}\else \v{r}\fi{}ezinov\'a}}, \bibinfo {author}
  {\bibfnamefont {T.}~\bibnamefont {Sato}}, \bibinfo {author} {\bibfnamefont
  {K.~L.}\ \bibnamefont {Ishikawa}}, \ and\ \bibinfo {author} {\bibfnamefont
  {J.}~\bibnamefont {Burgd\"orfer}},\ }\href {\doibase
  10.1103/PhysRevA.91.023412} {\bibfield  {journal} {\bibinfo  {journal} {Phys.
  Rev. A}\ }\textbf {\bibinfo {volume} {91}},\ \bibinfo {pages} {023412}
  (\bibinfo {year} {2015})}\BibitemShut {NoStop}%
\bibitem [{\citenamefont {Lackner}\ \emph {et~al.}(2017)\citenamefont
  {Lackner}, \citenamefont {B\ifmmode~\check{r}\else \v{r}\fi{}ezinov\'a},
  \citenamefont {Sato}, \citenamefont {Ishikawa},\ and\ \citenamefont
  {Burgd\"orfer}}]{PhysRevA.95.033414}%
  \BibitemOpen
  \bibfield  {author} {\bibinfo {author} {\bibfnamefont {F.}~\bibnamefont
  {Lackner}}, \bibinfo {author} {\bibfnamefont {I.}~\bibnamefont
  {B\ifmmode~\check{r}\else \v{r}\fi{}ezinov\'a}}, \bibinfo {author}
  {\bibfnamefont {T.}~\bibnamefont {Sato}}, \bibinfo {author} {\bibfnamefont
  {K.~L.}\ \bibnamefont {Ishikawa}}, \ and\ \bibinfo {author} {\bibfnamefont
  {J.}~\bibnamefont {Burgd\"orfer}},\ }\href {\doibase
  10.1103/PhysRevA.95.033414} {\bibfield  {journal} {\bibinfo  {journal} {Phys.
  Rev. A}\ }\textbf {\bibinfo {volume} {95}},\ \bibinfo {pages} {033414}
  (\bibinfo {year} {2017})}\BibitemShut {NoStop}%
\bibitem [{\citenamefont {Rohringer}\ \emph {et~al.}(2006)\citenamefont
  {Rohringer}, \citenamefont {Gordon},\ and\ \citenamefont
  {Santra}}]{Rohringer_TDCIS_2006}%
  \BibitemOpen
  \bibfield  {author} {\bibinfo {author} {\bibfnamefont {N.}~\bibnamefont
  {Rohringer}}, \bibinfo {author} {\bibfnamefont {A.}~\bibnamefont {Gordon}}, \
  and\ \bibinfo {author} {\bibfnamefont {R.}~\bibnamefont {Santra}},\ }\href
  {\doibase 10.1103/PhysRevA.74.043420} {\bibfield  {journal} {\bibinfo
  {journal} {Phys. Rev. A}\ }\textbf {\bibinfo {volume} {74}},\ \bibinfo
  {pages} {043420} (\bibinfo {year} {2006})}\BibitemShut {NoStop}%
\bibitem [{\citenamefont {Gordon}\ \emph {et~al.}(2006)\citenamefont {Gordon},
  \citenamefont {K\"artner}, \citenamefont {Rohringer},\ and\ \citenamefont
  {Santra}}]{PhysRevLett.96.223902}%
  \BibitemOpen
  \bibfield  {author} {\bibinfo {author} {\bibfnamefont {A.}~\bibnamefont
  {Gordon}}, \bibinfo {author} {\bibfnamefont {F.~X.}\ \bibnamefont
  {K\"artner}}, \bibinfo {author} {\bibfnamefont {N.}~\bibnamefont
  {Rohringer}}, \ and\ \bibinfo {author} {\bibfnamefont {R.}~\bibnamefont
  {Santra}},\ }\href {\doibase 10.1103/PhysRevLett.96.223902} {\bibfield
  {journal} {\bibinfo  {journal} {Phys. Rev. Lett.}\ }\textbf {\bibinfo
  {volume} {96}},\ \bibinfo {pages} {223902} (\bibinfo {year}
  {2006})}\BibitemShut {NoStop}%
\bibitem [{\citenamefont {Karamatskou}\ and\ \citenamefont
  {Santra}(2017)}]{Karam_CISXe_2017}%
  \BibitemOpen
  \bibfield  {author} {\bibinfo {author} {\bibfnamefont {A.}~\bibnamefont
  {Karamatskou}}\ and\ \bibinfo {author} {\bibfnamefont {R.}~\bibnamefont
  {Santra}},\ }\href {\doibase 10.1103/PhysRevA.95.013415} {\bibfield
  {journal} {\bibinfo  {journal} {Phys. Rev. A}\ }\textbf {\bibinfo {volume}
  {95}},\ \bibinfo {pages} {013415} (\bibinfo {year} {2017})}\BibitemShut
  {NoStop}%
\bibitem [{\citenamefont {Pabst}\ \emph {et~al.}(2011)\citenamefont {Pabst},
  \citenamefont {Greenman}, \citenamefont {Ho}, \citenamefont {Mazziotti},\
  and\ \citenamefont {Santra}}]{PhysRevLett.106.053003}%
  \BibitemOpen
  \bibfield  {author} {\bibinfo {author} {\bibfnamefont {S.}~\bibnamefont
  {Pabst}}, \bibinfo {author} {\bibfnamefont {L.}~\bibnamefont {Greenman}},
  \bibinfo {author} {\bibfnamefont {P.~J.}\ \bibnamefont {Ho}}, \bibinfo
  {author} {\bibfnamefont {D.~A.}\ \bibnamefont {Mazziotti}}, \ and\ \bibinfo
  {author} {\bibfnamefont {R.}~\bibnamefont {Santra}},\ }\href {\doibase
  10.1103/PhysRevLett.106.053003} {\bibfield  {journal} {\bibinfo  {journal}
  {Phys. Rev. Lett.}\ }\textbf {\bibinfo {volume} {106}},\ \bibinfo {pages}
  {053003} (\bibinfo {year} {2011})}\BibitemShut {NoStop}%
\bibitem [{\citenamefont {Greenman}\ \emph {et~al.}(2010)\citenamefont
  {Greenman}, \citenamefont {Ho}, \citenamefont {Pabst}, \citenamefont
  {Kamarchik}, \citenamefont {Mazziotti},\ and\ \citenamefont
  {Santra}}]{Greenman_TDCIS_2010}%
  \BibitemOpen
  \bibfield  {author} {\bibinfo {author} {\bibfnamefont {L.}~\bibnamefont
  {Greenman}}, \bibinfo {author} {\bibfnamefont {P.~J.}\ \bibnamefont {Ho}},
  \bibinfo {author} {\bibfnamefont {S.}~\bibnamefont {Pabst}}, \bibinfo
  {author} {\bibfnamefont {E.}~\bibnamefont {Kamarchik}}, \bibinfo {author}
  {\bibfnamefont {D.~A.}\ \bibnamefont {Mazziotti}}, \ and\ \bibinfo {author}
  {\bibfnamefont {R.}~\bibnamefont {Santra}},\ }\href {\doibase
  10.1103/PhysRevA.82.023406} {\bibfield  {journal} {\bibinfo  {journal} {Phys.
  Rev. A}\ }\textbf {\bibinfo {volume} {82}},\ \bibinfo {pages} {023406}
  (\bibinfo {year} {2010})}\BibitemShut {NoStop}%
\bibitem [{\citenamefont {Grosser}\ \emph {et~al.}(2017)\citenamefont
  {Grosser}, \citenamefont {Slowik},\ and\ \citenamefont
  {Santra}}]{PhysRevA.95.062107}%
  \BibitemOpen
  \bibfield  {author} {\bibinfo {author} {\bibfnamefont {M.}~\bibnamefont
  {Grosser}}, \bibinfo {author} {\bibfnamefont {J.~M.}\ \bibnamefont {Slowik}},
  \ and\ \bibinfo {author} {\bibfnamefont {R.}~\bibnamefont {Santra}},\ }\href
  {\doibase 10.1103/PhysRevA.95.062107} {\bibfield  {journal} {\bibinfo
  {journal} {Phys. Rev. A}\ }\textbf {\bibinfo {volume} {95}},\ \bibinfo
  {pages} {062107} (\bibinfo {year} {2017})}\BibitemShut {NoStop}%
\bibitem [{\citenamefont {Pabst}\ and\ \citenamefont
  {Santra}(2014)}]{Pabst_JPB_2014}%
  \BibitemOpen
  \bibfield  {author} {\bibinfo {author} {\bibfnamefont {S.}~\bibnamefont
  {Pabst}}\ and\ \bibinfo {author} {\bibfnamefont {R.}~\bibnamefont {Santra}},\
  }\href {http://stacks.iop.org/0953-4075/47/i=12/a=124026} {\bibfield
  {journal} {\bibinfo  {journal} {J. Phys. B: At., Mol. Opt. Phys.}\ }\textbf
  {\bibinfo {volume} {47}},\ \bibinfo {pages} {124026} (\bibinfo {year}
  {2014})}\BibitemShut {NoStop}%
\bibitem [{\citenamefont {Sytcheva}\ \emph {et~al.}(2012)\citenamefont
  {Sytcheva}, \citenamefont {Pabst}, \citenamefont {Son},\ and\ \citenamefont
  {Santra}}]{PhysRevA.85.023414}%
  \BibitemOpen
  \bibfield  {author} {\bibinfo {author} {\bibfnamefont {A.}~\bibnamefont
  {Sytcheva}}, \bibinfo {author} {\bibfnamefont {S.}~\bibnamefont {Pabst}},
  \bibinfo {author} {\bibfnamefont {S.-K.}\ \bibnamefont {Son}}, \ and\
  \bibinfo {author} {\bibfnamefont {R.}~\bibnamefont {Santra}},\ }\href
  {\doibase 10.1103/PhysRevA.85.023414} {\bibfield  {journal} {\bibinfo
  {journal} {Phys. Rev. A}\ }\textbf {\bibinfo {volume} {85}},\ \bibinfo
  {pages} {023414} (\bibinfo {year} {2012})}\BibitemShut {NoStop}%
\bibitem [{\citenamefont {Pabst}\ \emph {et~al.}(2012)\citenamefont {Pabst},
  \citenamefont {Sytcheva}, \citenamefont {Moulet}, \citenamefont {Wirth},
  \citenamefont {Goulielmakis},\ and\ \citenamefont
  {Santra}}]{PhysRevA.86.063411}%
  \BibitemOpen
  \bibfield  {author} {\bibinfo {author} {\bibfnamefont {S.}~\bibnamefont
  {Pabst}}, \bibinfo {author} {\bibfnamefont {A.}~\bibnamefont {Sytcheva}},
  \bibinfo {author} {\bibfnamefont {A.}~\bibnamefont {Moulet}}, \bibinfo
  {author} {\bibfnamefont {A.}~\bibnamefont {Wirth}}, \bibinfo {author}
  {\bibfnamefont {E.}~\bibnamefont {Goulielmakis}}, \ and\ \bibinfo {author}
  {\bibfnamefont {R.}~\bibnamefont {Santra}},\ }\href {\doibase
  10.1103/PhysRevA.86.063411} {\bibfield  {journal} {\bibinfo  {journal} {Phys.
  Rev. A}\ }\textbf {\bibinfo {volume} {86}},\ \bibinfo {pages} {063411}
  (\bibinfo {year} {2012})}\BibitemShut {NoStop}%
\bibitem [{\citenamefont {Pabst}\ and\ \citenamefont
  {Santra}(2013)}]{PhysRevLett.111.233005}%
  \BibitemOpen
  \bibfield  {author} {\bibinfo {author} {\bibfnamefont {S.}~\bibnamefont
  {Pabst}}\ and\ \bibinfo {author} {\bibfnamefont {R.}~\bibnamefont {Santra}},\
  }\href {\doibase 10.1103/PhysRevLett.111.233005} {\bibfield  {journal}
  {\bibinfo  {journal} {Phys. Rev. Lett.}\ }\textbf {\bibinfo {volume} {111}},\
  \bibinfo {pages} {233005} (\bibinfo {year} {2013})}\BibitemShut {NoStop}%
\bibitem [{\citenamefont {Heinrich-Josties}\ \emph {et~al.}(2014)\citenamefont
  {Heinrich-Josties}, \citenamefont {Pabst},\ and\ \citenamefont
  {Santra}}]{PhysRevA.89.043415}%
  \BibitemOpen
  \bibfield  {author} {\bibinfo {author} {\bibfnamefont {E.}~\bibnamefont
  {Heinrich-Josties}}, \bibinfo {author} {\bibfnamefont {S.}~\bibnamefont
  {Pabst}}, \ and\ \bibinfo {author} {\bibfnamefont {R.}~\bibnamefont
  {Santra}},\ }\href {\doibase 10.1103/PhysRevA.89.043415} {\bibfield
  {journal} {\bibinfo  {journal} {Phys. Rev. A}\ }\textbf {\bibinfo {volume}
  {89}},\ \bibinfo {pages} {043415} (\bibinfo {year} {2014})}\BibitemShut
  {NoStop}%
\bibitem [{\citenamefont {You}\ \emph {et~al.}(2017)\citenamefont {You},
  \citenamefont {Dahlstr\"om},\ and\ \citenamefont
  {Rohringer}}]{PhysRevA.95.023409}%
  \BibitemOpen
  \bibfield  {author} {\bibinfo {author} {\bibfnamefont {J.~A.}\ \bibnamefont
  {You}}, \bibinfo {author} {\bibfnamefont {J.~M.}\ \bibnamefont
  {Dahlstr\"om}}, \ and\ \bibinfo {author} {\bibfnamefont {N.}~\bibnamefont
  {Rohringer}},\ }\href {\doibase 10.1103/PhysRevA.95.023409} {\bibfield
  {journal} {\bibinfo  {journal} {Phys. Rev. A}\ }\textbf {\bibinfo {volume}
  {95}},\ \bibinfo {pages} {023409} (\bibinfo {year} {2017})}\BibitemShut
  {NoStop}%
\bibitem [{\citenamefont {You}\ \emph {et~al.}(2016)\citenamefont {You},
  \citenamefont {Rohringer},\ and\ \citenamefont
  {Dahlstr\"om}}]{PhysRevA.93.033413}%
  \BibitemOpen
  \bibfield  {author} {\bibinfo {author} {\bibfnamefont {J.~A.}\ \bibnamefont
  {You}}, \bibinfo {author} {\bibfnamefont {N.}~\bibnamefont {Rohringer}}, \
  and\ \bibinfo {author} {\bibfnamefont {J.~M.}\ \bibnamefont {Dahlstr\"om}},\
  }\href {\doibase 10.1103/PhysRevA.93.033413} {\bibfield  {journal} {\bibinfo
  {journal} {Phys. Rev. A}\ }\textbf {\bibinfo {volume} {93}},\ \bibinfo
  {pages} {033413} (\bibinfo {year} {2016})}\BibitemShut {NoStop}%
\bibitem [{\citenamefont {Sato}\ \emph
  {et~al.}(2018{\natexlab{b}})\citenamefont {Sato}, \citenamefont {Teramura},\
  and\ \citenamefont {Ishikawa}}]{Sato_Gauge_2018}%
  \BibitemOpen
  \bibfield  {author} {\bibinfo {author} {\bibfnamefont {T.}~\bibnamefont
  {Sato}}, \bibinfo {author} {\bibfnamefont {T.}~\bibnamefont {Teramura}}, \
  and\ \bibinfo {author} {\bibfnamefont {K.~L.}\ \bibnamefont {Ishikawa}},\
  }\href {\doibase 10.3390/app8030433} {\bibfield  {journal} {\bibinfo
  {journal} {Appl. Sci}\ }\textbf {\bibinfo {volume} {8}},\ \bibinfo {pages}
  {433} (\bibinfo {year} {2018}{\natexlab{b}})}\BibitemShut {NoStop}%
\bibitem [{\citenamefont {Rescigno}\ and\ \citenamefont
  {McCurdy}(2000)}]{PhysRevA.62.032706}%
  \BibitemOpen
  \bibfield  {author} {\bibinfo {author} {\bibfnamefont {T.~N.}\ \bibnamefont
  {Rescigno}}\ and\ \bibinfo {author} {\bibfnamefont {C.~W.}\ \bibnamefont
  {McCurdy}},\ }\href {\doibase 10.1103/PhysRevA.62.032706} {\bibfield
  {journal} {\bibinfo  {journal} {Phys. Rev. A}\ }\textbf {\bibinfo {volume}
  {62}},\ \bibinfo {pages} {032706} (\bibinfo {year} {2000})}\BibitemShut
  {NoStop}%
\bibitem [{\citenamefont {McCurdy}\ \emph {et~al.}(2004)\citenamefont
  {McCurdy}, \citenamefont {Baertschy},\ and\ \citenamefont
  {Rescigno}}]{McCurdy_2004}%
  \BibitemOpen
  \bibfield  {author} {\bibinfo {author} {\bibfnamefont {C.~W.}\ \bibnamefont
  {McCurdy}}, \bibinfo {author} {\bibfnamefont {M.}~\bibnamefont {Baertschy}},
  \ and\ \bibinfo {author} {\bibfnamefont {T.~N.}\ \bibnamefont {Rescigno}},\
  }\href {\doibase 10.1088/0953-4075/37/17/r01} {\bibfield  {journal} {\bibinfo
   {journal} {J. Phys. B: At., Mol. Opt. Phys.}\ }\textbf {\bibinfo {volume}
  {37}},\ \bibinfo {pages} {R137} (\bibinfo {year} {2004})}\BibitemShut
  {NoStop}%
\bibitem [{\citenamefont {Schneider}\ \emph {et~al.}(2006)\citenamefont
  {Schneider}, \citenamefont {Collins},\ and\ \citenamefont
  {Hu}}]{PhysRevE.73.036708}%
  \BibitemOpen
  \bibfield  {author} {\bibinfo {author} {\bibfnamefont {B.~I.}\ \bibnamefont
  {Schneider}}, \bibinfo {author} {\bibfnamefont {L.~A.}\ \bibnamefont
  {Collins}}, \ and\ \bibinfo {author} {\bibfnamefont {S.~X.}\ \bibnamefont
  {Hu}},\ }\href {\doibase 10.1103/PhysRevE.73.036708} {\bibfield  {journal}
  {\bibinfo  {journal} {Phys. Rev. E}\ }\textbf {\bibinfo {volume} {73}},\
  \bibinfo {pages} {036708} (\bibinfo {year} {2006})}\BibitemShut {NoStop}%
\bibitem [{\citenamefont {Schneider}\ \emph {et~al.}(2011)\citenamefont
  {Schneider}, \citenamefont {Johannes}, \citenamefont {Nagele}, \citenamefont
  {Pazourek}, \citenamefont {Hu}, \citenamefont {Collins},\ and\ \citenamefont
  {Burgd{\"{o}}rfer}}]{Schneider2011}%
  \BibitemOpen
  \bibfield  {author} {\bibinfo {author} {\bibfnamefont {B.~I.}\ \bibnamefont
  {Schneider}}, \bibinfo {author} {\bibfnamefont {F.}~\bibnamefont {Johannes}},
  \bibinfo {author} {\bibfnamefont {S.}~\bibnamefont {Nagele}}, \bibinfo
  {author} {\bibfnamefont {R.}~\bibnamefont {Pazourek}}, \bibinfo {author}
  {\bibfnamefont {S.}~\bibnamefont {Hu}}, \bibinfo {author} {\bibfnamefont
  {L.~A.}\ \bibnamefont {Collins}}, \ and\ \bibinfo {author} {\bibfnamefont
  {J.}~\bibnamefont {Burgd{\"{o}}rfer}},\ }\href {\doibase
  10.1007/978-1-4419-9491-2} {\emph {\bibinfo {title} {{Quantum Dynamic
  Imaging}}}},\ edited by\ \bibinfo {editor} {\bibfnamefont {A.~D.}\
  \bibnamefont {Bandrauk}}\ and\ \bibinfo {editor} {\bibfnamefont
  {M.}~\bibnamefont {Ivanov}}\ (\bibinfo  {publisher} {Springer New York},\
  \bibinfo {address} {New York, NY},\ \bibinfo {year} {2011})\ p.\ \bibinfo
  {pages} {149}\BibitemShut {NoStop}%
\bibitem [{\citenamefont {Kramer}\ and\ \citenamefont
  {Saraceno}(1981)}]{Kramer:107648}%
  \BibitemOpen
  \bibfield  {author} {\bibinfo {author} {\bibfnamefont {P.}~\bibnamefont
  {Kramer}}\ and\ \bibinfo {author} {\bibfnamefont {M.}~\bibnamefont
  {Saraceno}},\ }\href@noop {} {\emph {\bibinfo {title} {{Geometry of the
  time-dependent variational principle in quantum mechanics}}}}\ (\bibinfo
  {publisher} {Springer},\ \bibinfo {year} {1981})\BibitemShut {NoStop}%
\bibitem [{\citenamefont {Ishikawa}\ and\ \citenamefont
  {Sato}(2015)}]{Ishikawa2015}%
  \BibitemOpen
  \bibfield  {author} {\bibinfo {author} {\bibfnamefont {K.~L.}\ \bibnamefont
  {Ishikawa}}\ and\ \bibinfo {author} {\bibfnamefont {T.}~\bibnamefont
  {Sato}},\ }\href {\doibase 10.1109/JSTQE.2015.2438827} {\bibfield  {journal}
  {\bibinfo  {journal} {IEEE J. Sel. Top. Quantum Electron.}\ }\textbf
  {\bibinfo {volume} {21}} (\bibinfo {year} {2015}),\
  10.1109/JSTQE.2015.2438827}\BibitemShut {NoStop}%
\bibitem [{\citenamefont {Sato}\ \emph
  {et~al.}(2018{\natexlab{c}})\citenamefont {Sato}, \citenamefont {Orimo},
  \citenamefont {Teramura}, \citenamefont {Oyunbileg},\ and\ \citenamefont
  {Ishikawa}}]{PUILS}%
  \BibitemOpen
  \bibfield  {author} {\bibinfo {author} {\bibfnamefont {T.}~\bibnamefont
  {Sato}}, \bibinfo {author} {\bibfnamefont {Y.}~\bibnamefont {Orimo}},
  \bibinfo {author} {\bibfnamefont {T.}~\bibnamefont {Teramura}}, \bibinfo
  {author} {\bibfnamefont {T.}~\bibnamefont {Oyunbileg}}, \ and\ \bibinfo
  {author} {\bibfnamefont {K.~L.}\ \bibnamefont {Ishikawa}},\ }\href {\doibase
  10.1007/978-3-030-03786-4} {\emph {\bibinfo {title} {{Progress in Ultrafast
  Intense Laser Science XIV}}}},\ edited by\ \bibinfo {editor} {\bibfnamefont
  {K.}~\bibnamefont {Yamanouchi}}, \bibinfo {editor} {\bibfnamefont
  {P.}~\bibnamefont {Martin}}, \bibinfo {editor} {\bibfnamefont
  {M.}~\bibnamefont {Sentis}}, \bibinfo {editor} {\bibfnamefont
  {L.}~\bibnamefont {Ruxin}}, \ and\ \bibinfo {editor} {\bibfnamefont
  {D.}~\bibnamefont {Normand}},\ \bibinfo {series} {Springer Series in Chemical
  Physics}, Vol.\ \bibinfo {volume} {118}\ (\bibinfo  {publisher} {Springer
  International Publishing},\ \bibinfo {address} {Cham},\ \bibinfo {year}
  {2018})\ pp.\ \bibinfo {pages} {143--172}\BibitemShut {NoStop}%
\bibitem [{\citenamefont {Krogstad}(2005)}]{Krogstad2005}%
  \BibitemOpen
  \bibfield  {author} {\bibinfo {author} {\bibfnamefont {S.}~\bibnamefont
  {Krogstad}},\ }\href {\doibase 10.1016/j.jcp.2004.08.006} {\bibfield
  {journal} {\bibinfo  {journal} {J. Comput. Phys.}\ }\textbf {\bibinfo
  {volume} {203}},\ \bibinfo {pages} {72} (\bibinfo {year} {2005})}\BibitemShut
  {NoStop}%
\bibitem [{\citenamefont {Hochbruck}\ and\ \citenamefont
  {Ostermann}(2010)}]{hochbruck_ostermann_2010}%
  \BibitemOpen
  \bibfield  {author} {\bibinfo {author} {\bibfnamefont {M.}~\bibnamefont
  {Hochbruck}}\ and\ \bibinfo {author} {\bibfnamefont {A.}~\bibnamefont
  {Ostermann}},\ }\href {\doibase 10.1017/S0962492910000048} {\bibfield
  {journal} {\bibinfo  {journal} {Acta Numerica}\ }\textbf {\bibinfo {volume}
  {19}},\ \bibinfo {pages} {209–286} (\bibinfo {year} {2010})}\BibitemShut
  {NoStop}%
\bibitem [{\citenamefont {Kidd}\ \emph {et~al.}(2017)\citenamefont {Kidd},
  \citenamefont {Covington},\ and\ \citenamefont {Varga}}]{Kidd2017}%
  \BibitemOpen
  \bibfield  {author} {\bibinfo {author} {\bibfnamefont {D.}~\bibnamefont
  {Kidd}}, \bibinfo {author} {\bibfnamefont {C.}~\bibnamefont {Covington}}, \
  and\ \bibinfo {author} {\bibfnamefont {K.}~\bibnamefont {Varga}},\ }\href
  {\doibase 10.1103/PhysRevE.96.063307} {\bibfield  {journal} {\bibinfo
  {journal} {Phys. Rev. E}\ }\textbf {\bibinfo {volume} {96}},\ \bibinfo
  {pages} {1} (\bibinfo {year} {2017})}\BibitemShut {NoStop}%
\bibitem [{\citenamefont {Feist}\ \emph {et~al.}(2009)\citenamefont {Feist},
  \citenamefont {Nagele}, \citenamefont {Pazourek}, \citenamefont {Persson},
  \citenamefont {Schneider}, \citenamefont {Collins},\ and\ \citenamefont
  {Burgd\"orfer}}]{PhysRevLett.103.063002}%
  \BibitemOpen
  \bibfield  {author} {\bibinfo {author} {\bibfnamefont {J.}~\bibnamefont
  {Feist}}, \bibinfo {author} {\bibfnamefont {S.}~\bibnamefont {Nagele}},
  \bibinfo {author} {\bibfnamefont {R.}~\bibnamefont {Pazourek}}, \bibinfo
  {author} {\bibfnamefont {E.}~\bibnamefont {Persson}}, \bibinfo {author}
  {\bibfnamefont {B.~I.}\ \bibnamefont {Schneider}}, \bibinfo {author}
  {\bibfnamefont {L.~A.}\ \bibnamefont {Collins}}, \ and\ \bibinfo {author}
  {\bibfnamefont {J.}~\bibnamefont {Burgd\"orfer}},\ }\href {\doibase
  10.1103/PhysRevLett.103.063002} {\bibfield  {journal} {\bibinfo  {journal}
  {Phys. Rev. Lett.}\ }\textbf {\bibinfo {volume} {103}},\ \bibinfo {pages}
  {063002} (\bibinfo {year} {2009})}\BibitemShut {NoStop}%
\bibitem [{\citenamefont {Pazourek}\ \emph {et~al.}(2011)\citenamefont
  {Pazourek}, \citenamefont {Feist}, \citenamefont {Nagele}, \citenamefont
  {Persson}, \citenamefont {Schneider}, \citenamefont {Collins},\ and\
  \citenamefont {Burgd\"orfer}}]{PhysRevA.83.053418}%
  \BibitemOpen
  \bibfield  {author} {\bibinfo {author} {\bibfnamefont {R.}~\bibnamefont
  {Pazourek}}, \bibinfo {author} {\bibfnamefont {J.}~\bibnamefont {Feist}},
  \bibinfo {author} {\bibfnamefont {S.}~\bibnamefont {Nagele}}, \bibinfo
  {author} {\bibfnamefont {E.}~\bibnamefont {Persson}}, \bibinfo {author}
  {\bibfnamefont {B.~I.}\ \bibnamefont {Schneider}}, \bibinfo {author}
  {\bibfnamefont {L.~A.}\ \bibnamefont {Collins}}, \ and\ \bibinfo {author}
  {\bibfnamefont {J.}~\bibnamefont {Burgd\"orfer}},\ }\href {\doibase
  10.1103/PhysRevA.83.053418} {\bibfield  {journal} {\bibinfo  {journal} {Phys.
  Rev. A}\ }\textbf {\bibinfo {volume} {83}},\ \bibinfo {pages} {053418}
  (\bibinfo {year} {2011})}\BibitemShut {NoStop}%
\bibitem [{\citenamefont {Corkum}(1993)}]{PhysRevLett.71.1994}%
  \BibitemOpen
  \bibfield  {author} {\bibinfo {author} {\bibfnamefont {P.~B.}\ \bibnamefont
  {Corkum}},\ }\href {\doibase 10.1103/PhysRevLett.71.1994} {\bibfield
  {journal} {\bibinfo  {journal} {Phys. Rev. Lett.}\ }\textbf {\bibinfo
  {volume} {71}},\ \bibinfo {pages} {1994} (\bibinfo {year}
  {1993})}\BibitemShut {NoStop}%
\bibitem [{\citenamefont {Piraux}\ \emph {et~al.}(1993)\citenamefont {Piraux},
  \citenamefont {L'Huillier},\ and\ \citenamefont
  {Rz{\c{a}}{\.{z}}ewski}}]{Piraux1993}%
  \BibitemOpen
  \bibinfo {editor} {\bibfnamefont {B.}~\bibnamefont {Piraux}}, \bibinfo
  {editor} {\bibfnamefont {A.}~\bibnamefont {L'Huillier}}, \ and\ \bibinfo
  {editor} {\bibfnamefont {K.}~\bibnamefont {Rz{\c{a}}{\.{z}}ewski}},\ eds.,\
  \href {\doibase 10.1007/978-1-4615-7963-2} {\emph {\bibinfo {title}
  {{Super-Intense Laser-Atom Physics}}}},\ \bibinfo {series} {NATO ASI Series},
  Vol.\ \bibinfo {volume} {316}\ (\bibinfo  {publisher} {Springer US},\
  \bibinfo {address} {Boston, MA},\ \bibinfo {year} {1993})\ p.~\bibinfo
  {pages} {95}\BibitemShut {NoStop}%
\end{thebibliography}
